\documentclass[12pt,onecolumn,showpacs,amssymb,aps,nofootinbib,floatfix]{revtex4-1}
\usepackage{epsfig,color,bigints}
\usepackage{tikz-cd}

\newcommand{\de}[1]{\left(#1 \right)}
\newcommand{\abs}[1]{\left|#1 \right|}
\newcommand{\mean}[1]{\left<#1 \right>}

\newcommand{\De}[1]{\left[#1 \right]}
\newcommand{\DE}[1]{\left\{#1 \right\}}

\newcommand{\coss}[2]{\text{cos}^{#2}{\de{#1}}}
\newcommand{\ket}[1]{\left\vert #1 \right>}
\newcommand{\bra}[1]{\left< #1 \right\vert}
\newcommand{\DD}[2]{\frac{\delta #1}{\delta #2}}

\newcommand{\splt}[1]{\begin{equation}\begin{split} #1 \end{split} \end{equation}}
\newcommand{\inner}[2]{\left<#1\vert#2 \right>}

\newcommand{\ave}[1]{\left\langle #1 \right\rangle}

\renewcommand{\eqref}[1]{Eq. \ref{#1} }

\newcommand{\eqcomma}{\phantom{AA},\phantom{AA}}

\begin{document}
\title{The Hanbury-Brown and Twiss effect in inflationary cosmological perturbations}
\author{Gustavo Matheus Gauy, Flavia Sobreira, Giorgio Torrieri}
\affiliation{Universidade Estadual de Campinas - Instituto de Fisica "Gleb Wataghin"\\
Rua Sérgio Buarque de Holanda, 777\\
 CEP 13083-859 - Campinas SP\\
}
\begin{abstract}
  The simplest model of inflation is based around an inflaton field that starts in a coherent false vacuum state with a positive cosmological constant, rolls slowly to the true vacuum and relaxes to it via reheating.  We examine whether the scale of the transition from coherence to chaoticity can be examined via the Hanbury-Brown and Twiss (HBT) effect, in parallel with analogous problems of heavy ion physics (the ``pion laser'' and the thermalizing glasma).
  We develop an ansatz which contains a definition of ''chaoticity'' which parallels that of the usual setups where HBT is used.
  However, we also discuss the differences between the inflationary setup and more mainstream uses of HBT and conclude that these are more significant than the similarities, making the use of the developed methodology uncertain.
\end{abstract}
\maketitle

\section{Introduction}
Inflation is the leading paradigm to understand the early evolution of the universe \cite{Baumann2022}.  The current universe's homogeneity, flatness and large-scale structure can be thought of arising out of an ``inflaton'' field which transitions from a coherent false vacuum with a positive cosmological constant to the true vacuum it currently occupies via a ``roll'' semiclassical evolution followed by rapid oscillations quenched by interactions with standard model particles (``reheating'').   The quantum fluctuations around this classical evolution are the simplest explanation of the observed perturbations in the Cosmic Microwave background.
If the roll is sufficiently slow fluctuations can be assumed to be both Gaussian and scale-invariant \cite{Baumann2022}.

However, this picture is still largely theoretical.  We do not know the nature of the inflation and the details, or even the qualitative scales, of the stages of the above evolution, bar a very rough estimate.

In this work, we will examine the question of whether the HBT effect, long used as a source to understand issues of coherence and configuration space length scales in heavy ion collisions \cite{Glauber1963,Glauber1963a,Glauber2007,gyulassy,heinz,lisa,Sinyukov2013}, can be used to probe these issues further.
We note that the Hanburgy-Brown Twiss effect has been has been proposed before as a probe \cite{soda1,soda2,Giovannini2010,Giovannini2011,Giovannini2017,Giovannini2017a,Giovannini2019} but not of the Inflaton field and not in the manner explored in this work.   Our main motivation is that the general dynamics of inflation-reheating has some similarity with the Glasma thermalization of a heavy ion collision \cite{glasma}, and the latter has been examined via HBT correlations \cite{kovner}.


While a detailed introduction of the HBT effect, including a disquisition over its relationship with fundamental quantum mechanics, is left for the appendix, we shall comment that its basic idea is a direct consequence of Bose-Einstein statistics.  A classical coherent field producing spin 0 bosons is governed by the Klein-Gordon equation with a source
\begin{equation}
\left( \partial^2 +m^2 +V\left(\phi(x),x \right) \right) \phi(x)=J(x) \Leftrightarrow \left( k^2 + m^2 + F\left(\left\{ \tilde{V}_k,\tilde{\phi}_k \right\} \right) \right) \tilde{\phi}_k=J_k
  \end{equation}
For a coherent source, $J_k$s are all in the same phase.   If, however, interactions between $\phi_k$s are strong and non-linear, a de-phasing occurs.  In the limit of local thermalization \cite{gyulassy,heinz,lisa} one can model this by giving $J_k$ a ``random'' phase, $J_k \rightarrow \exp[i \hat{\phi}_k] J_k$, where
$\hat{\phi}$ is a random operator.

Such random phase is a picture of both Statistical mechanics \cite{huang} and quantum-chaotic evolution \cite{berry}, and it is in fact easy to check that in the flat potential limit the onset of quantum chaos results in the same random phase \cite{srednicki} that is used to characterize maximally incoherent sources in HBT \cite{Glauber2007,gyulassy}.

For the limit of an extended source (a star, or a heavy ion collision) emitting on-shell free bosons which then reach a pair of detectors, the coherent and chaotic limits can be distinguished by 2-particle correlations (or equivalently $\ave{a^{+}_{k} a_{k} a^{+}_{k'}a_{k'}}$, the creation and annihillation operators), the probability of finding two particles with momenta $k_{1,2}$ w.r.t. a randomized background.
This will be
\begin{equation}
  \label{hbtdef}
  \frac{P(k_1,k_2)}{P(k_1) P(k_2)} \propto 1+\alpha \frac{\left|\int d^4 x S(x,k_1+k_2)e^{i(k_1-k_2).x} \right|^{2}}{\int d^4 x S(x,k_1) \int d^4 x S(x,k_2)} 
  \eqcomma{\alpha=
    \begin{array}{cc}
      0 & \mathrm{coherent}\\
      1 & \mathrm{chaotic}
      \end{array}
}  \end{equation}
Where $S(x,k)$ is a classical probability density function to emit a particle of momentum $k$ from a point of position $x$.
In the intermediate regime, between coherent and chaotic, $\alpha$ will be a function of $k_1 \pm k_2$.   In astronomy this has been used as a method to measure the size of stars (its original post-war application \cite{Brown1954} which had a recent resurgence \cite{lisastars,lisastars2,lisastars3}), in hadronic collisions it is likewise used as a probe of the spacetime scale of the system \cite{lisa}, as well as a probe of coherence of the underlying fields \cite{laser}.

Hence, in principle the equivalent of the HBT effect in inflation could be used to understand its structure in ''configuration space'' (ie the causal scale of the universe when reheating occurs and its duration) as well as the change in coherence during the reheating.   The equivalent of \eqref{hbtdef} in the absence of an on-shell inflaton would be the ''trispectrum'', the 4-point function of curvature perturbations $\ave{\phi(x_1)\phi(x_2)\phi(x_3)\phi(x_4)}$. The set-up related to astronomy and heavy ion physics, with two sources and two detectors, is really a 4-point function with two sides stretched to large space distances and respectively, in the past for the sources and the future for the detectors.  In the rest of this work we shall assess the conceptual and technical appropriateness of this idea.

We note that the effect of disorder in inflation has been discussed previously \cite{disorder1,disorder2}, with some discussion and even quantitative conclusions (the effect of the trispectrum) somewhat similar to ours.  
However, as discussed in detail in the appendix, the HBT phenomenon is on appearence semiclassical but relies on two deeply ''quantum'' characteristics, indistiguishability of identical particles and phase coherence.   While in \cite{disorder1,disorder2} fluctuations are generated by the quantum nature of the inflaton, they are trated by classical Fokker-Planck dynamics, and hence, as explicitly discussed in these works, the effect survives (and is in fact more easily calculable) in the limit of many inflaton fields.   This is an important point since there is evidence that in both quantum mechanics \cite{srednicki} and field theory \cite{mrow} quantum phases randomize on a much faster scale than momenta and in fact are the main driver of equilibration.   Thus, where \cite{disorder1,disorder2} need complicated multi-field potentials (analogous to solids with impurities), a generic non-linear potential together with high occupation numbers (where the Boltzmann limit is valid \cite{muller}) is required for the randomization inherent in the maximally incoherent source used in HBT.
Thus, in the spirit of this fast phase randomization produced by generic chaotic potentials suggested by \cite{berry,srednicki,mrow} we explore an ansatz where random phases are introduced ''by hand'', and examine what effect this has on correlations which are as similar as possible to HBT. 
\section{An HBT like ansatz \label{secansatz}}
Not a lot is known on how the process of reheating affects primordial curvature perturbations. A crucial feature of these perturbations is they are constant on superhorizon scales. In fact, one can show that is true up to all loop corrections when these perturbations are generated by the quantum fluctuations of a single inflaton field\cite{Senatore2012,Assassi2012}. This could lead one to conclude there couldn't be an observable effect of reheating over superhorizon modes and, therefore, no consequence to the formation of cosmological structure. But that's not necessarily the case. During reheating, the inflaton will have to decay into fields of the standard model otherwise, without this transfer of energy, the universe would be empty of standard model particles after inflation. The point is these interactions of the inflaton with other fields will come in non-linear corrections and could result in non-constant superhorizon modes even if these interactions are primarily on subhorizon scales. So, it's possible the process of reheating could have some effect over the correlations of the perturbations generated by the inflaton \cite{brand1,brand2,brand3}.  

The quantum fluctuations of the inflaton classicalize upon horizon exit, hence loop corrections due to interactions of the inflaton with other fields can be thought of as quantum corrections over these classical configurations. The usual assumption is, then, that these corrections should be small and perturbation theory is employed. Under this assumption, observable effect should only be a small correction over the standard prediction of inflation. However, it's possible this isn't the case for interactions during reheating. The coupling between the inflaton and other fields should be dressed by the growing mode functions of the inflaton\cite{Kaya2013}. If reheating lasts long enough, these coupling could become very large. This would lead the inflaton to exhibit strong mode-mode coupling and highly non-linear behavior.  If inflation is ``warm'' \cite{warm} such chaoticization effects could be particularly prominent, since they would last in the whole inflationary era.
As it is known, strongly coupled harmonic oscillators behave in a very chaotic way\cite{Easther1997,Cornish1996}. So, we expect the same to happen with the inflaton modes: they should exhibit some chaotic behavior. This is the assumption we will be working with in this work.

	There is some parallel with the interplay of a coherent field into a thermalized gas which was thought to occur in chiral condensates \cite{laser,gyulassy} as well as in the ``Glasma'' \cite{glasma}.  In both these cases it was assumed that the thermalized part is chaotic, and the signal would be encoded in the coherence of HBT-type identical particle correlations.
	
	In this work, we will attempt to model this effect reheating could have on superhorizon modes. The idea is to propose an \emph{ansatz} to take the conjectured chaoticity of the modes into account. Then, look for constraints imposed by observations and if, even after that, there are any new predictions.

	\subsection{Chaotic ansatz}
	
	There are two ways in which chance could come into the theory: out of the initial conditions or the dynamics. In the standard lore of inflation, the initial conditions for inflationary perturbations are random, since they originate from quantum fluctuations, but their time evolution is linear and deterministic. Reheating could introduce a new source of randomness of the dynamical kind. As argued before, interactions of the inflaton with other fields during reheating could lead to some chaoticity to be developed over the superhorizon modes. 
	It's unlikely one could show that analytically. 
	So, we attempt to introduce this effect by proposing an \emph{ansatz}. 
	
	Comoving curvature perturbations can be expanded, up to first order in the slow-roll parameters, as
	\splt{
		\mathcal{R}(\textbf{x},\tau) 
		= 
		\int d^3k e^{i\textbf{k} \cdot \textbf{x}}\De{  g\de{k,\tau} a\de{ \textbf{k} } + g^*\de{k,\tau} a^\dagger\de{- \textbf{k} } };
	}
	where $g\de{k,\tau} =\frac{f\de{k,\tau}}{z(\tau) \de{2\pi}^3 \sqrt{2k}}$. The above expression is for the free theory, without taking into account the interactions during reheating. These perturbations are solutions of the free classical field equation and have very specified trajectories in phase-space. We will call them coherent perturbations. A chaotic or incoherent field, on the other hand, won't have well determined trajectories in phase-space. They are the result of very non-linear dynamics, developing a chaotic behavior to their trajectories. We attempt to model that with the following \emph{ansatz}:
	\splt{
		g(k,\tau) \mapsto \gamma(k,\tau)g(k,\tau);
		\label{an}
		}
	where $\gamma\de{k,\tau}$ is some complex Gaussian random function. Meaning, the trajectories are, now, determined by the classical equations of motion and an ensemble of the kind:
	\splt{
		\mean{F\de{\gamma,\gamma^*} }_\gamma
		= 
		\int \mathcal{D}\gamma e^{- \gamma^* \cdot M^{-1} \cdot \gamma} F\de{\gamma, \gamma^*}
		, \qquad
		\int \mathcal{D}\gamma e^{- \gamma^* \cdot M^{-1} \cdot \gamma } = 1.
	}
	Because it's Gaussian, the ensemble is completely determined by the 2-point function $\mean{\gamma^*(k,\tau) \gamma(p,\tau')} = M(p,k,\tau',\tau)$.
	
	Given the \emph{ansatz} (\ref{an}), it's straightforward to calculate the chaotic equal time correlation functions. 
	Now, the idea is to see which constraints observations impose over the \emph{ansatz}. Since  \cite{Baumann2022} the fluctuations in the temperature of the CMB and the large scale structure of the early universe can be directly connected to the perturbations produced during inflation. As we saw, they indicate these fluctuations are Gaussian random functions with an almost scale invariant spectrum. 
	Starting from the lowest order, it's simple to see the \emph{ansatz} won't change the 1-point correlation function.

	For the 2-point function or spectrum, we get
    For the 2-point function or spectrum, we get
        \splt{
                \mean{\mathcal{R}_\gamma\de{\textbf{x},\tau} \mathcal{R}_\gamma\de{\textbf{y},\tau} }
                &=
                \mean{ \bra{0}\mathcal{R}_\gamma\de{\textbf{x},\tau} \mathcal{R}_\gamma\de{\textbf{y},\tau} \ket{0} }_{\gamma}
                \\
                &=
                \mean{K_\gamma\de{\textbf{x} - \textbf{y},\tau}}_\gamma
                \\
                &=
                \int d^3k e^{i \textbf{k}\cdot\de{\textbf{x} - \textbf{y}}} \mean{\abs{\gamma(k,\tau)}^2}_{\gamma} \abs{g\de{k,\tau}}^2.
        }

	Since  \cite{Baumann2022} the spectrum obtained with coherent perturbations, as we have defined them, has the precise qualitative properties to explain observations, since they result in an almost scale-invariant power spectrum for superhorizon scales. Our \emph{ansatz} is potentially adding a new source of momentum or scale dependence. 
	Hence, in order to be compatible with observations, one must assume for superhorizon scales:
	\splt{
		\mean{\abs{\gamma(k,\tau)}^2}_\gamma = 1.
	}
	The other option would be to evaluate an unequal time correlation function and see what changes could result from a chaotic perturbation. That has already been considered in the context of perturbations due to cosmic strings\cite{Magueijo1995,Magueijo1996,Albrecht1995}. The behavior of chaotic perturbations would be the same as active perturbations in the string situation\cite{Bassett1998}. That is, incoherence in the time component of perturbations lead to the destruction of secondary acoustic peaks. So, this possibility has already been discarded by observations, the perturbations should be coherent in the time component at least up to the scales observed so far. A fully chaotic spectrum in the time component is incompatible with that. This implies the constraint
	\splt{
		M(k,p,\tau',\tau) = M(k,p), \text{ or } \gamma(k,\tau) = \gamma(k)
		}
	over the ensemble. 
	
	We haven't been able to find any new prediction of the ansatz up to second order correlation functions. Hence, let's look at higher order. 
	The equal time 3-point correlation function or bispectrum will also remain unchanged by symmetry, in accordance with observations\cite{Orlando2023}.

        Thus the first non-trivial result will come from the equal time 4-point correlation function or trispectrum, which is given by:
	\splt{
		\mean{\mathcal{R}_\gamma\de{\textbf{x},\tau}\mathcal{R}_\gamma\de{\textbf{y},\tau}\mathcal{R}_\gamma\de{\textbf{w},\tau}\mathcal{R}_\gamma\de{\textbf{z},\tau}}
		&=
		\mean{ K_\gamma\de{ \textbf{x} - \textbf{y},\tau } K_\gamma\de{ \textbf{w} - \textbf{z},\tau } }_{\gamma} 
		\\
		&+
		\mean{ K_\gamma\de{ \textbf{x} - \textbf{w},\tau } K_\gamma\de{ \textbf{y} - \textbf{z},\tau } }_{\gamma}
		\\
		&+
		\mean{ K_\gamma\de{ \textbf{x} - \textbf{z},\tau } K_\gamma\de{ \textbf{y} - \textbf{w} ,\tau} }_{\gamma};
	}
	could still me modified non-trivially by the \emph{ansatz}, as long as $M(k,p,\tau, \tau)$ is not a constant for $k\not = p$. To see that, all one has to do is calculate explicitly the ensemble average:
	\splt{
		\mean{ K_\gamma\de{ \textbf{x} - \textbf{y},\tau } K_\gamma\de{ \textbf{w} - \textbf{z},\tau } }_\gamma
		&=
		\int d^3 k d^3 p e^{i\textbf{k}\cdot\de{\textbf{x} - \textbf{y} }} e^{i\textbf{k}\cdot\de{\textbf{w} - \textbf{z} }} \mean{\abs{\gamma(k,\tau)}^2\abs{\gamma(p,\tau)}^2}_\gamma \abs{g(k,\tau)}^2 \abs{g(p,\tau)}^2
		\\
		&=\int d^3 k d^3 p
	 e^{i\textbf{k}\cdot\de{\textbf{x} - \textbf{y} }} e^{i\textbf{k}\cdot\de{\textbf{w} - \textbf{z} }} \abs{g(k,\tau)}^2 \abs{g(p,\tau)}^2
		\\
		&\times \de{\mean{\abs{\gamma(k,\tau)}^2}_\gamma\mean{\abs{\gamma(p,\tau)}^2}_\gamma +  \abs{ \mean{\gamma^*(k,\tau)\gamma(p,\tau) }}^2 }
	}
	So, after imposing the constraint obtained from the power spectrum, it will be given by
	\splt{
		\mean{ K_\gamma\de{ \textbf{x} - \textbf{y},\tau } K_\gamma\de{ \textbf{w} - \textbf{z},\tau } }_\gamma
		&=
		K(\textbf{x} - \textbf{y},\tau) K(\textbf{w} - \textbf{z},\tau)
		\\
		&+
		\int d^3 k d^3 p e^{i\textbf{k}\cdot\de{\textbf{x} - \textbf{y} }} e^{i\textbf{k}\cdot\de{\textbf{w} - \textbf{z} }} \abs{g(k,\tau)}^2 \abs{g(p,\tau)}^2
		\abs{ \mean{\gamma^*(k,\tau)\gamma(p,\tau) }}^2  
	}
	where the first term is the component already present in the coherent trispectrum (via the Wightman functions $K(x-y,\tau) \equiv K_{\gamma=1}$ ,see \eqref{wightdef} for the Minkowski case) and the second one is the correction predicted by the ansatz. Then, with the previous result, the chaotic trispectrum is given by
	\splt{
		&\mean{\mathcal{R}_\gamma\de{\textbf{x},\tau}\mathcal{R}_\gamma\de{\textbf{y},\tau}\mathcal{R}_\gamma\de{\textbf{w},\tau}\mathcal{R}_\gamma\de{\textbf{z},\tau}}
		=
		\bra{0}\mathcal{R}\de{\textbf{x},\tau}\mathcal{R}\de{\textbf{y},\tau}\mathcal{R}\de{\textbf{w},\tau}\mathcal{R}\de{\textbf{z},\tau}\ket{0}
		\\
		&+
		\int d^3 k d^3 p\de{ 
			e^{i\textbf{k}\cdot\de{\textbf{x} - \textbf{y} }}
			e^{i\textbf{k}\cdot\de{\textbf{w} - \textbf{z} }}
			+
			e^{i\textbf{k}\cdot\de{\textbf{x} - \textbf{w} }} 
			e^{i\textbf{k}\cdot\de{\textbf{y} - \textbf{z} }} 
			+
			e^{i\textbf{k}\cdot\de{\textbf{x} - \textbf{z} }} 
			e^{i\textbf{k}\cdot\de{\textbf{y} - \textbf{w} }}
		}\times
		\\
		&\times\abs{g(k,\tau)}^2 \abs{g(p,\tau)}^2
		\abs{ \mean{\gamma^*(k,\tau)\gamma(p,\tau) }}^2,
	}
	where $\mathcal{R}$ denotes coherent perturbations. 
	Because of the assumption of a Gaussian ensemble, there are no new predictions for unequal time correlation functions beyond the constraints imposed by the power spectrum. Meaning, there is no need to look into the unequal time 4-point correlation function. 
	Considering both constraints imposed by the power spectrum over the ensemble and the central limit theorem, one possible choice for $M$ is
	\splt{
		M\de{k,p} = e^{-\frac{\tau^2_{\text{rh}}}{2} \de{k - p}^2 },
	}		
	where $\tau_{\text{rh}}$ is some time scale related to reheating. Therefore, under these assumptions the trispectrum would be modified to
	\splt{
		&\mean{\mathcal{R}_\gamma\de{\textbf{x},\tau}\mathcal{R}_\gamma\de{\textbf{y},\tau}\mathcal{R}_\gamma\de{\textbf{w},\tau}\mathcal{R}_\gamma\de{\textbf{z},\tau}}_\gamma
		=
		\bra{0}\mathcal{R}\de{\textbf{x},\tau}\mathcal{R}\de{\textbf{y},\tau}\mathcal{R}\de{\textbf{w},\tau}\mathcal{R}\de{\textbf{z},\tau}\ket{0}
		\\
		&+
			\int_{\tau_{r}^{1}}^\Lambda k^2 d k d\Omega_k \int_{\tau_{r}^{1}}^\Lambda p^2 dp d\Omega_p \de{ 
			e^{i\textbf{k}\cdot\de{\textbf{x} - \textbf{y} }}
			e^{i\textbf{k}\cdot\de{\textbf{w} - \textbf{z} }}
			+
			e^{i\textbf{k}\cdot\de{\textbf{x} - \textbf{w} }} 
			e^{i\textbf{k}\cdot\de{\textbf{y} - \textbf{z} }} 
			+
			e^{i\textbf{k}\cdot\de{\textbf{x} - \textbf{z} }} 
			e^{i\textbf{k}\cdot\de{\textbf{y} - \textbf{w} }}
		}
                        \times	}
        \[\  \times	\abs{g(k,\tau)}^2 \abs{g(p,\tau)}^2
		e^{-\tau^2_{\text{rh}} \de{k - p}^2 }.  \]
	This is not an analytic integral, but it is finite since it has physical higher and lower limits, .   The lower limit will be the infrared $1/\tau_{r}$ corresponding to the horizon scale at reheating.   The upper limit $\Lambda$ will be the scale at which the phase randomizes (the ``coherence domain'' of the inflaton.  In the usual HBT \cite{gyulassy,heinz,lisa} this cutoff is not necessary because of on-shellness of the interfering quanta).   If this model is to be developed phenomenologically, trispectrum features need to be fitted to $\tau_r,\Lambda$, this would be a subject of an evnetual further work.
	
	\subsection{Partially coherent perturbations}
	
	One can define partially chaotic perturbations by the same \emph{ansatz} with just a different ensemble. On this situation, the ensemble will have a coherent component defined by the 1-point function
	\splt{
		\mean{\gamma(k,\tau)}_\gamma = \gamma_0(k,\tau) \not = 0,
		}
	So, to impose a partially chaotic ensemble, all one must do is to assume that the ensemble is defined by
	\splt{
		\mean{\cdots}_\gamma 
		= 
		\int \mathcal{D}\gamma \exp\De{ - \de{\gamma - \gamma_0}^* \cdot M^{-1}\cdot \de{\gamma - \gamma_0}  }\de{\cdots}.
		}	
	Given the above ensemble, we must calculate averages such as
	\splt{
		\mean{\gamma^*(k_1,\tau_1) \gamma(k_2,\tau_2)}_\gamma, \ \mean{ \gamma^*(k_1,\tau_1) \gamma^*(k_2,\tau_2) \gamma(k_3,\tau_3)\gamma(k_4, \tau_4) }_\gamma, \ \cdots
		} 
	Hence, in order to do that, we work out the generating functional for such terms:
	\splt{
		Z(f,f^*)
		&=
		\mean{
		e^{\gamma^* \cdot  f}e^{\gamma\cdot f^*}
		}_\gamma
		=
		\int \mathcal{D}\gamma \exp\De{ - \de{\gamma - \gamma_0}^* \cdot M^{-1}\cdot \de{\gamma - \gamma_0}  } e^{\gamma^* \cdot  f}e^{\gamma\cdot f^*}
		\\
		&=
		\int \mathcal{D}\gamma \exp\De{ - \gamma^*\cdot M^{-1}\cdot \gamma^*  } e^{\de{\gamma + \gamma_0}^* \cdot  f}e^{\de{\gamma+ \gamma_0}\cdot f^*}
		\\
		&=
		\exp\de{\gamma_0 \cdot f^* + \gamma^*_0\cdot f + f^*\cdot M \cdot f } Z(0),
		}
	where $Z(0)$ is just a normalization. Therefore, the generating functional can be rewritten as
	\splt{
		Z(f,f^*) = e^{W\de{f,f^*}} Z(0),
		}
	where all the information is stored in		
	\splt{
		W = \gamma_0 \cdot f^* + \gamma^*_0\cdot f + f^*\cdot M \cdot f,
		}
	and the normalization $Z(0)$ can be taken to be one. 
	
	With the generating functional $Z(f,f^*)$, we can finally calculate the $\gamma$ ensemble averages. We only need two terms: the 2-point function
	\splt{
		\mean{\gamma^*(k,\tau) \gamma(p,\tau')}_\gamma,
		}
	and the four point function
	\splt{
		\mean{\abs{\gamma(k,\tau)}^2 \abs{\gamma(p,\tau')}^2},
		} 
	since these are the only terms that will be relevant for the calculations of the spectrum and trispectrum of cosmological perturbations.		
	For the 2-point function
	\splt{
		\mean{\gamma^*(k,\tau)\gamma(p,\tau')}_\gamma
		&=
		\frac{\delta^2 Z}{\delta f(k,\tau) \delta f^*(p,\tau')}(0)
		\\
		&=
		\frac{\delta^2 W}{\delta f(k,\tau) \delta f^*(p,\tau')}(0) + \frac{\delta W}{\delta f(k,\tau)}(0)\frac{\delta W}{\delta f^*(p,\tau')}(0), 
		}
	resulting in
	\splt{
	 	\mean{\gamma^*(k,\tau)\gamma(p,\tau')}_{\gamma} 
	 	= 
	 	M(p,k,\tau',\tau) + \gamma_0(p,\tau')\gamma^*_0(k,\tau).	
	}
	While the four-point function in question gives
	\splt{
		\mean{\abs{\gamma(k,\tau)}^2 \abs{\gamma(p,\tau')}^2  }_\gamma
		&=
		\frac{\delta^4 Z}{\delta f(k)f(p)f^*(p)f^*(k)} (0)
		\\
		&=
		\frac{\delta^2 W}{\delta f(k) \delta f^*(k)}(0)
		\frac{\delta^2 W}{\delta f(p) \delta f^*(p)}(0)
		+
		\frac{\delta^2 W}{\delta f(k) \delta f^*(p)}(0)
		\frac{\delta^2 W}{\delta f(p) \delta f^*(k)}(0)
		\\
		&+
		\frac{\delta^2 W}{\delta f(k) \delta f^*(k)}(0) 
		\abs{\frac{\delta W}{\delta f^*(p)}(0)}^2
		+
		\frac{\delta^2 W}{\delta f(p) \delta f^*(p)}(0) 
		\abs{\frac{\delta W}{\delta f^*(k)}(0)}^2
		\\
		&+
		\frac{\delta^2 W}{\delta f(p) \delta f^*(k)}(0) 
		\frac{\delta W}{\delta f^*(p)}(0)\frac{\delta W}{\delta f(k)}(0)
		+
		\frac{\delta^2 W}{\delta f(k) \delta f^*(p)}(0) 
		\frac{\delta W}{\delta f^*(k)}(0)\frac{\delta W}{\delta f(p)}(0)
		\\
		&+
		\abs{\frac{\delta W}{\delta f^*(k)}(0)}^2\abs{\frac{\delta W}{\delta f^*(p)}(0)}^2.
		}
	Therefore, this 4-point function can be written in terms of the 2-point and 1-point functions as:
	\splt{
		\mean{\abs{\gamma(k)}^2 \abs{\gamma(p)}^2 }_\gamma 
		=
		\mean{\abs{\gamma(k)}^2}_\gamma\mean{\abs{\gamma(p)}^2}_\gamma
		+
		\abs{\mean{\gamma^*(k)\gamma(p)}_\gamma}^2
		-
		\abs{\mean{\gamma(k)}_\gamma}^2 \abs{\mean{\gamma(p)}_\gamma}^2.
		}
	With these results, we have everything needed to calculate the correlation functions of partially chaotic cosmological perturbations.
	
	As observations tell us, the perturbations are random and we know their spectrum. So, again, we must impose these constraints. The first one, is that perturbations are random:
	\splt{
		\mean{\mathcal{R}_\gamma} = 0.
		}
	That is immediately true, even for a partially chaotic ensemble. With this choice of \emph{ansatz}, this is true for any ensemble since 
	\splt{
		\bra{0}a_{\textbf{k}}\ket{0} = \bra{0} a^\dagger_{\textbf{k}}\ket{0} = 0.
		} 
	This means there are no constraints imposed at this order. Similarly to the fully chaotic situation, the constraints will come from the second order. The equal time 2-point correlation function once again imposes
	\splt{
		\mean{\abs{\gamma(k,\tau)}^2} = 1,
		}	
	in order for the \emph{ansatz} not to be in conflict with observations. The level of coherence is not determined at this order. But, this time, we can't immediately disconsider the time dependence of the ensemble. It's possible for the perturbations to be coherent on the time variable at large scales, the ones we observe, but incoherent on smaller scales. That would have implications for the spectrum, resulting in the lack of multiple peak structure in those smaller scales, but preserving it on higher scales. This would be a new prediction, but is yet to be observed. So, for simplicity, we will assume to be working only with high enough scales in order to impose, once again, the constraint:
	\splt{
		\gamma(k,\tau) = \gamma(k).
		}
	With this assumption, new predictions are only possible on higher orders.  Once again, the bispectrum is not affected as shown from symmetry considerations.
	
 The trispectrum, on the other hand, will have some new implications. As we know from the previous section, the trispectrum can be expressed as
	\splt{
		\mean{\mathcal{R}_\gamma\de{\textbf{x},\tau}\mathcal{R}_\gamma\de{\textbf{y},\tau}\mathcal{R}_\gamma\de{\textbf{w},\tau}\mathcal{R}_\gamma\de{\textbf{z},\tau}}
		&=
		\mean{ K_\gamma\de{ \textbf{x} - \textbf{y},\tau } K_\gamma\de{ \textbf{w} - \textbf{z},\tau } }_{\gamma} 
		\\
		&+
		\mean{ K_\gamma\de{ \textbf{x} - \textbf{w},\tau } K_\gamma\de{ \textbf{y} - \textbf{z},\tau } }_{\gamma}
		\\
		&+
		\mean{ K_\gamma\de{ \textbf{x} - \textbf{z},\tau } K_\gamma\de{ \textbf{y} - \textbf{w} ,\tau} }_{\gamma}.
	}
	Considering a partially chaotic ensemble, one gets
	\splt{
		&\mean{ K_\gamma\de{ \textbf{x} - \textbf{y},\tau } K_\gamma\de{ \textbf{w} - \textbf{z},\tau } }_\gamma
		=
		\int d^3 k d^3 p e^{i\textbf{k}\cdot\de{\textbf{x} - \textbf{y} }} e^{i\textbf{k}\cdot\de{\textbf{w} - \textbf{z} }} \mean{\abs{\gamma(k,\tau)}^2\abs{\gamma(p,\tau)}^2}_\gamma \abs{g(k,\tau)}^2 \abs{g(p,\tau)}^2
		\\
		&=
		\int d^3 k d^3 p e^{i\textbf{k}\cdot\de{\textbf{x} - \textbf{y} }} e^{i\textbf{k}\cdot\de{\textbf{w} - \textbf{z} }} \abs{g(k,\tau)}^2 \abs{g(p,\tau)}^2
		\\
		&\times \de{
			\mean{\abs{\gamma(k,\tau)}^2}\mean{\abs{\gamma(p,\tau)}^2} + \abs{\mean{\gamma^*(k,\tau)\gamma(p,\tau)}}^2 - \abs{\mean{\gamma(k,\tau)}}^2 \abs{\mean{\gamma(p,\tau)}}^2
			}.
		}
	With the constraints imposed by the spectrum, one can simplify the trispectrum to
	\splt{
		&\mean{ K_\gamma\de{ \textbf{x} - \textbf{y},\tau } K_\gamma\de{ \textbf{w} - \textbf{z},\tau } }_\gamma
		=
		K\de{ \textbf{x} - \textbf{y},\tau } K\de{ \textbf{w} - \textbf{z},\tau }
		\\
		&+
		\int d^3 k d^3 p e^{i\textbf{k}\cdot\de{\textbf{x} - \textbf{y} }} e^{i\textbf{k}\cdot\de{\textbf{w} - \textbf{z} }} \abs{g(k,\tau)}^2 \abs{g(p,\tau)}^2
		\de{
			 \abs{\mean{\gamma^*(k)\gamma(p)}}^2 - \abs{\mean{\gamma(k)}}^2 \abs{\mean{\gamma(p)}}^2
		},
	}	
	where the second term is the correction due to the partially chaotic ensemble. Since there is a coherent component, a choice for the ensemble, that is compatible with the spectrum constraints, is a little more involved in this situation. Following \cite{Boal1990}, we will assume it can be written as
	\splt{
		\mean{\gamma^*(k)\gamma(p)}_\gamma
		=
		\gamma^*_0(k)\gamma_0(p) + \gamma^*_{\text{ch}}(k)\gamma_{\text{ch}}(p)\rho(k- p),
		}
	where $\rho$ is a real number and $\rho(0) = 1$. Hence, the spectrum constraints imply
	\splt{
		\mean{\abs{\gamma(k)}^2} = 1 \iff \abs{\gamma_0(k)}^2 + \abs{\gamma_{ch}(k)}^2\ = 1.
		}	
	With this assumption, we can express
	\splt{
		&\abs{\mean{\gamma^*(k)\gamma(p)}}^2 - \abs{\mean{\gamma(k)}}^2 \abs{\mean{\gamma(p)}}^2
		=
		\abs{\gamma_{ch}(k)}^2\abs{\gamma_{ch}(p)}^2\rho^2(k-p) 
		\\
		&+ 
		\de{ 
			\gamma_0^*(k)\gamma_0(p) \gamma_{\text{ch}}^*(p)\gamma_{\text{ch}}(k)
			+
			\gamma_0(k)\gamma^*_0(p) \gamma_{\text{ch}}(p)\gamma^*_{\text{ch}}(k)
			 }
		\rho(k-p).	  
		}
	Assuming $\gamma_0$ and $\gamma_{\text{ch}}$ are real functions, one can rewrite
	\splt{
		\abs{\mean{\gamma^*(k)\gamma(p)}}^2 - \abs{\mean{\gamma(k)}}^2 \abs{\mean{\gamma(p)}}^2
		&=
		2\gamma_0(k)\gamma_0(p) \sqrt{1 - \gamma^2_0(k)}\sqrt{1 - \gamma^2_0(p)} \rho(k - p) 
		\\
		&+
		\de{1 - \gamma^2_0(k)}\de{1 - \gamma^2_0(p)} \rho^2(k - p).
		}
	Defining the degree of coherence as
	\splt{
		D(k) = \abs{\gamma_0(k)}^2.
	}
	With this definition, one gets
	\splt{
		&\mean{ K_\gamma\de{ \textbf{x} - \textbf{y},\tau } K_\gamma\de{ \textbf{w} - \textbf{z},\tau } }_\gamma
		=
		K\de{ \textbf{x} - \textbf{y},\tau } K\de{ \textbf{w} - \textbf{z},\tau }
		\\
		&+
		\int d^3 k d^3 p e^{i\textbf{k}\cdot\de{\textbf{x} - \textbf{y} }} e^{i\textbf{k}\cdot\de{\textbf{w} - \textbf{z} }} \abs{g(k,\tau)}^2 \abs{g(p,\tau)}^2
		\\
		&\times\De{
			2\sqrt{
				D(k)D(p)\de{1-D(k)}\de{1 - D(p)}
				}
				\rho(k-p)
			+
			\de{1- D(k)}\de{1-D(p)} \rho^2(k-p)	
		}.
	}
	Assuming again the chaotic part is described by a Gaussian:
	\splt{
		\rho(k - p) = e^{-\frac{\tau^2_{\text{rh}}}{2}\de{k - p}^2};
		}
	the partially chaotic trispectrum can, therefore, be written as
	\splt{
		&\mean{\mathcal{R}_\gamma\de{\textbf{x},\tau}\mathcal{R}_\gamma\de{\textbf{y},\tau}\mathcal{R}_\gamma\de{\textbf{w},\tau}\mathcal{R}_\gamma\de{\textbf{z},\tau}}_\gamma
		=
		\bra{0}\mathcal{R}\de{\textbf{x},\tau}\mathcal{R}\de{\textbf{y},\tau}\mathcal{R}\de{\textbf{w},\tau}\mathcal{R}\de{\textbf{z},\tau}\ket{0}
		\\
		&+
		\int d^3 k d^3 p\de{ 
			e^{i\textbf{k}\cdot\de{\textbf{x} - \textbf{y} }}
			e^{i\textbf{k}\cdot\de{\textbf{w} - \textbf{z} }}
			+
			e^{i\textbf{k}\cdot\de{\textbf{x} - \textbf{w} }} 
			e^{i\textbf{k}\cdot\de{\textbf{y} - \textbf{z} }} 
			+
			e^{i\textbf{k}\cdot\de{\textbf{x} - \textbf{z} }} 
			e^{i\textbf{k}\cdot\de{\textbf{y} - \textbf{w} }}
		}
		\abs{g(k,\tau)}^2 \abs{g(p,\tau)}^2
		\\
		&\times
		\De{
			2\sqrt{
				D(k)D(p)\de{1-D(k)}\de{1 - D(p)}
			}
			e^{-\frac{\tau^2_{\text{rh}}}{2}\de{k - p}^2}
			+
			\de{1- D(k)}\de{1-D(p)} e^{-\tau^2_{\text{rh}}\de{k - p}^2}
		}.
	}
	From the above expression, we see that for $D(k) = 1$ full coherence is recovered, while for $D(k) = 0$ one gets the chaotic case. The above formalism allows a few possibilities. Every scales could be partially chaotic, that is the degree of coherence being independent of scale: $D(k) = D$. Or only some scales could exhibit chaoticity. For example $D(k) = \Theta(k - k_0)$, where only scales larger than some $k_0$ would be chaotic. This is once again not an integral one can do analytically. Given a choice for the degree of coherence and cutoffs for the integral, one needs to numerically integrate it in order to compare with observations.

	\section{ Discussion: Is this really an HBT effect? } 
	\subsection{Virtuality and coherence of inflaton perturbations}
	Initially we thought there would be some connection between the formalism described in \cite{Glauber1963} with the \emph{ansatz} and results obtained in the last section. There are two perspectives one could take for the HBT effect, we will call them HBT correlations and Bose-Einstein correlations. In the first one, the HBT effect is just the presence of correlations in the measurements of two particle detectors. So, for this approach, one must build an HBT interferometer for inflatons. The first problem one faces when trying to study inflationary perturbations from this perspective is that we don't have inflaton detectors. Every measurement we supposedly make of the inflaton is an indirect one, where the inflaton is in a sense {\em virtual}, a 2-point correlator in configuration space, classicalized by evolution.  In addition, the measurement erases any time-ordering (it is not two sources and two detectors but rather a 4-point function).  There is no obvious relationship between the space-like 4-point function in the sky related to the trispectrum and the 4-point function connecting sources in the infinite past to detectors in the infinite future relevant for interferometry of on-shell photons and pions.

        Still, we did find a non-Gaussianity in the trispectrum which explicitly depends on the chaoticity.  Is this HBT?     A further problem with interpreting our result this way is, the inflaton is not coherent in the HBT sense. So it's meaningless to say it gained some chaoticity or incoherence from reheating, at least in the HBT sense. To prove it isn't coherent, we only need to show the second order correlation function doesn't factorize. For inflaton perturbations, due to the expansion of spacetime, the evolution operator acts as a Bogolyubov transformation
	\splt{
		a\de{\textbf{k},\tau} 
		&= 
		U^\dagger(-\infty,\tau) a\de{\textbf{k}}U(-\infty,\tau) 
		\\
		&=
		\alpha(k,\tau) a\de{\textbf{k}} + \beta^*(k,\tau)a^\dagger\de{-\textbf{k}}.
		}
	The correlation function for two detectors, at time $\tau$, results in
	\splt{
		\mean{a^\dagger\de{\textbf{k}}a^\dagger\de{\textbf{p}}a\de{\textbf{p}}a\de{\textbf{k}}}(\tau)
		&=
		\bra{0}U^\dagger(-\infty,\tau)a^\dagger\de{\textbf{k}}a^\dagger\de{\textbf{p}}a\de{\textbf{p}}a\de{\textbf{k}} U(-\infty,\tau)\ket{0}
		\\
		&=
		\bra{0}U^\dagger(-\infty,\tau) a^\dagger\de{\textbf{k}}U(-\infty,\tau)U^\dagger(-\infty,\tau) a^\dagger\de{\textbf{p}}U(-\infty,\tau)
		\\
		&\times U^\dagger(-\infty,\tau) a\de{\textbf{p}}U(-\infty,\tau)U^\dagger(-\infty,\tau) a\de{\textbf{k}}U(-\infty,\tau)\ket{0}
		\\
		&=
		\bra{0}a^\dagger\de{\textbf{k},\tau}a^\dagger\de{\textbf{p},\tau}a\de{\textbf{p},\tau}a\de{\textbf{k},\tau}\ket{0}. 
		}
	Then, applying the Bogolyubov transformation, one gets after some algebra
	\splt{
		\mean{a^\dagger\de{\textbf{k}}a^\dagger\de{\textbf{p}}a\de{\textbf{p}}a\de{\textbf{k}}}(\tau)
		&=
		\abs{\beta\de{k,\tau}}^2\abs{\beta\de{p,\tau}}^2 \bra{0}a\de{-\textbf{k}}a\de{-\textbf{p}}a^\dagger\de{-\textbf{p}}a^\dagger\de{-\textbf{k}}\ket{0} 
		\\
		&+
		\abs{\beta\de{k,\tau}}\abs{\alpha\de{p,\tau}}^2 
		\bra{0}a\de{-\textbf{k}} a^\dagger\de{\textbf{p}} a\de{\textbf{p}}a^\dagger\de{-\textbf{k}}\ket{0}.
		}
	To show the non-factorization of the second order correlation function, all we need to do is normal order the above expected values. The first term results in
	\splt{
		\bra{0}a\de{-\textbf{k}}a\de{-\textbf{p}}a^\dagger\de{-\textbf{p}}a^\dagger\de{-\textbf{k}}\ket{0} 
		&=
		\de{2\pi}^6\de{\delta(0)\delta(0) + \delta\de{\textbf{k} - \textbf{p}}\delta\de{\textbf{k} - \textbf{p}}},
		}
	while the last one
	\splt{
		\bra{0}a\de{-\textbf{k}}a^\dagger\de{\textbf{p}} a\de{\textbf{p}}a^\dagger\de{-\textbf{k}}\ket{0}
		&=
		\de{2\pi}^6\delta\de{\textbf{k}+ \textbf{p}} \delta\de{\textbf{k}+ \textbf{p}} .
		}
	Now, using the relations
	\splt{
		\mean{a^\dagger\de{\textbf{k}} a\de{\textbf{p}} }\de{\tau}
		&=
		\de{2\pi}^3 \beta\de{k,\tau}\beta^*\de{p,\tau} \delta\de{\textbf{k} - \textbf{p}},
		}	
	\splt{
		\mean{a\de{\textbf{k}} a\de{\textbf{p}} }\de{\tau}
		&=
		\de{2\pi}^3\alpha\de{k,\tau}\beta^*\de{p,\tau}\delta\de{\textbf{k} + \textbf{p}},
		}	
	we can rewrite the correlation function in question as
	\splt{
		\mean{a^\dagger\de{\textbf{k}}a^\dagger\de{\textbf{p}}a\de{\textbf{p}}a\de{\textbf{k}}}(\tau)
		&=
		\mean{a^\dagger\de{\textbf{k}}a\de{\textbf{k}}}\de{\tau}
		\mean{a^\dagger\de{\textbf{p}}a\de{\textbf{p}}}\de{\tau}
		\\
		&+
		\abs{\mean{a^\dagger\de{\textbf{k}}a\de{\textbf{p}}}\de{\tau}}^2
		+
		\abs{\mean{a\de{\textbf{k}}a\de{\textbf{k}}}\de{\tau}}^2.
		}
	Hence, as claimed, the second order correlation function doesn't factorize:
	\splt{
		\mean{a^\dagger\de{\textbf{k}}a^\dagger\de{\textbf{p}}a\de{\textbf{p}}a\de{\textbf{k}}}(\tau)
		\not=
		\mean{a^\dagger\de{\textbf{k}}a\de{\textbf{k}}}\de{\tau}
		\mean{a^\dagger\de{\textbf{p}}a\de{\textbf{p}}}\de{\tau}.
		}
	By following the exact same procedure, one can also show that
	\splt{
		\mean{\varphi^\dagger\de{\textbf{x}} \varphi^\dagger\de{ \textbf{y} } \varphi\de{\textbf{y}} \varphi\de{\textbf{x}} }\de{\tau}
		&=
		\mean{\varphi^\dagger\de{\textbf{x}}\varphi\de{\textbf{x}} }\de{\tau}
		\mean{\varphi^\dagger\de{\textbf{y}}\varphi\de{\textbf{y}} }\de{\tau} 
		\\
		&+
		\abs{\mean{\varphi^\dagger\de{\textbf{x}}\varphi\de{\textbf{y}} }\de{\tau} }^2  
		+
		\abs{\mean{\varphi\de{\textbf{x}}\varphi\de{\textbf{y}} }\de{\tau} }^2,  
		} 
	where
	\splt{ 
		\varphi(\textbf{x},\tau) 
		&=
		\int \frac{d^3k}{\de{2\pi}^3} e^{i\textbf{k}\cdot \textbf{x}} 
		\de{
			\alpha\de{k,\tau}a\de{\textbf{k}} + \beta^*\de{k,\tau}a^\dagger\de{-\textbf{k}}
			}.
	}
	Meaning, there would be a correlation, at the time $\tau$, between two detectors at positions $\textbf{x}$ and $\textbf{y}$:
	\splt{
		\frac{\mean{\varphi^\dagger\de{\textbf{x}} \varphi^\dagger\de{ \textbf{y} } \varphi\de{\textbf{y}} \varphi\de{\textbf{x}} }\de{\tau}}
		{\mean{\varphi^\dagger\de{\textbf{x}}\varphi\de{\textbf{x}} }\de{\tau}
		\mean{\varphi^\dagger\de{\textbf{y}}\varphi\de{\textbf{y}} }\de{\tau} }
		=
		1 
		+ 
		\frac{
		\abs{\mean{\varphi^\dagger\de{\textbf{x}}\varphi\de{\textbf{y}} }\de{\tau} }^2  
		+
		\abs{\mean{\varphi\de{\textbf{x}}\varphi\de{\textbf{y}} }\de{\tau} }^2}
		{
			\mean{\varphi^\dagger\de{\textbf{x}}\varphi\de{\textbf{x}} }\de{\tau}
			\mean{\varphi^\dagger\de{\textbf{y}}\varphi\de{\textbf{y}} }\de{\tau}
			}
		}
	and, therefore, the HBT effect, in the sense of HBT correlations, for primordial perturbations, without any need for reheating to introduce chaoticity.   Physically, the expansion of spacetime ''changes the vacuum'' of spacetime, and this is equivalent, in a quantum field theory, to changing the definition of the coherent state.  In heavy ion physics this decoherence by change in the vacuum state was actually already proposed as a probe of chiral symmetry restoration \cite{gyulchiral}.

        This means that the chaoticity proposed in the \emph{ansatz} is not necessarily related to the one used for external classical sources in standard HBT. As formulated by Glauber\cite{Glauber1963}, incoherence is defined by the presence of HBT correlations in an intensity interferometer. Therefore, under the standard quantum optics formalism, inflaton perturbations are already incoherent or chaotic, and an $\alpha>0$ (\eqref{hbtdef}) will be present also before reheating and a quantitative model-dependent analysis will have to be used to isolate each contribution. 
	\subsection{The role of Bose-Einstein statistics}
	The other perspective one could approach this problems is based on Bose-Einstein correlations. We thought there wouldn't be Bose-Einstein correlations after horizon crossing, since the inflaton classicalizes and classical particles are distinguishable. There are a number of problems with this idea. There is no such thing as a classical particle in quantum field theory, what classicalizes is the field itself. But, nonetheless, one could expect Bose-Einstein correlations to disappear in the classical limit. Unfortunately, that is also not true. We show that now. Consider a state with two sets of particles: $n$ particles with momentum $\textbf{k}$ and another set of $m$ particles with momentum $\textbf{p}$, assume $\textbf{k}\not = \textbf{p}$. That is, the state
	\splt{
		\ket{n_{\textbf{k}}, m_{\textbf{p}}} = \frac{1}{\sqrt{n!}}\de{\frac{a^\dagger(\textbf{k}) }{\sqrt{\de{2\pi}^3\delta(0)}} }^n
		\frac{1}{\sqrt{m!}}\de{\frac{a^\dagger(\textbf{p}) }{\sqrt{\de{2\pi}^3\delta(0)}} }^m
		\ket{0}
		}
	where each creation operator carries a factor $\frac{1}{\sqrt{\de{2\pi}^3\delta(0)} }$ for normalization purposes, $\delta(0)$ should just be interpreted as the volume of space. The amplitude for two particles to be absorbed by two detectors at $\textbf{x}$ and $\textbf{y}$ is:
	\splt{
		&
		\varphi(\textbf{x}) \varphi(\textbf{y})\ket{n_{\textbf{k}}, m_{\textbf{p}}}
		=
		\sqrt{nm}\frac{e^{i\textbf{k}\cdot\textbf{x}}}{\sqrt{2k}}
		\frac{e^{i\textbf{p}\cdot\textbf{y}}}{\sqrt{2p}}
		\de{\frac{1}{\sqrt{\de{2\pi}^3 \delta(0)}}}^{n+m}  
		\frac{\de{ a^\dagger(\textbf{k}) }^{n-1}}{ \sqrt{\de{n-1}!} }
		\frac{\de{ a^\dagger(\textbf{p}) }^{m-1}}{\sqrt{\de{m-1}!}}
		\ket{0}
		\\
		&+
		\sqrt{nm}\frac{e^{i\textbf{p}\cdot\textbf{x}}}{\sqrt{2p}}
		\frac{e^{i\textbf{k}\cdot\textbf{y}}}{\sqrt{2k}}
		\de{\frac{1}{\sqrt{\de{2\pi}^3 \delta(0)}}}^{n+m}  
		\frac{\de{ a^\dagger(\textbf{k}) }^{n-1}}{ \sqrt{\de{n-1}!} }
		\frac{\de{ a^\dagger(\textbf{p}) }^{m-1}}{\sqrt{\de{m-1}!}}
		\ket{0}
		\\
		&+
		\sqrt{n\de{n-1}}\frac{e^{i\textbf{k}\cdot\textbf{x}}}{\sqrt{2k}}
		\frac{e^{i\textbf{k}\cdot\textbf{y}}}{\sqrt{2k}}
		\de{\frac{1}{\sqrt{\de{2\pi}^3 \delta(0)}}}^{n+m}  
		\frac{\de{ a^\dagger(\textbf{k}) }^{n-2}}{ \sqrt{\de{n-2}!} }
		\frac{\de{ a^\dagger(\textbf{p}) }^{m}}{\sqrt{m!}}
		\ket{0}
		\\
		&+
		\sqrt{m\de{m-1}}\frac{e^{i\textbf{p}\cdot\textbf{x}}}{\sqrt{2p}}
		\frac{e^{i\textbf{p}\cdot\textbf{y}}}{\sqrt{2p}}
		\de{\frac{1}{\sqrt{\de{2\pi}^3 \delta(0)}}}^{n+m}  
		\frac{\de{ a^\dagger(\textbf{k}) }^{n}}{ \sqrt{n!} }
		\frac{\de{ a^\dagger(\textbf{p}) }^{m-2}}{\sqrt{\de{m-2}!}}
		\ket{0},
		}	
	where we have used:
	\splt{
		\De{\varphi(\textbf{x}), \de{a^\dagger(\textbf{k})}^n}
		&=
		\De{\int \frac{d^3q}{\sqrt{2q}\de{2\pi}^3} e^{i\textbf{q}\cdot \textbf{x}} a\de{\textbf{q}}, \de{a^\dagger(\textbf{k})}^n }
		=
		n \frac{e^{i\textbf{k}\cdot\textbf{x}}}{\sqrt{2k}} \de{a^\dagger\de{\textbf{k}}}^{n-1}.
		}	
	As an example, consider the situation when $n=m=1$:
	\splt{
		&\varphi(\textbf{x}) \varphi(\textbf{y})
		\ket{1_{\textbf{k}},1_{\textbf{p}}}
		=
		\frac{	
		e^{i\textbf{k}\cdot\textbf{x}}
		e^{i\textbf{p}\cdot\textbf{y}}
		+
		e^{i\textbf{p}\cdot\textbf{x}}
		e^{i\textbf{k}\cdot\textbf{y}}
		}{2 \sqrt{kp}}
		\frac{1}{\de{2\pi}^3\delta(0)}\ket{0}.
		}
	So, as expected, we see the Bose-Einstein symmetry in the above amplitude. Going back to the general case, the probability of simultaneous detections is then
	\splt{
		&\bra{n_{\textbf{k}},m_{\textbf{p}}} 
		\varphi^\dagger(\textbf{y}) \varphi^\dagger(\textbf{x})
		\varphi(\textbf{x}) \varphi(\textbf{y})
		\ket{n_\textbf{k},m_{\textbf{p}}}
		=
		\frac{nm}{4kp}\abs{e^{i\textbf{k}\cdot \textbf{x}} e^{i\textbf{p}\cdot \textbf{y}} + e^{i\textbf{p}\cdot \textbf{x}} e^{i\textbf{k}\cdot \textbf{y}} }^2 \de{\frac{1}{\de{2\pi}^3 \delta(0)}}^{2}
		\\
		&+
		\frac{n\de{n-1}}{4k^2}\de{\frac{1}{\de{2\pi}^3 \delta(0)}}^{2}
		+
		\frac{m\de{m-1}}{4p^2}\de{\frac{1}{\de{2\pi}^3 \delta(0)}}^{2}
		\\
		&=
		\de{\frac{1}{\de{2\pi}^3 2\delta(0)}}^{2}
		\De{
			2\frac{nm}{kp}\de{1 + \coss{\de{\textbf{k}-\textbf{p}} \de{\textbf{x} -\textbf{y} } }{}}
			+
			\frac{n\de{n-1}}{k^2}
			+
			\frac{m\de{m-1}}{p^2}
			}.
		}
	If $n=m=1$, we get
	\splt{
		\bra{1_{\textbf{k}},1_{\textbf{p}}} 
		\varphi^\dagger(\textbf{y}) \varphi^\dagger(\textbf{x})
		\varphi(\textbf{x}) \varphi(\textbf{y})
		\ket{1_\textbf{k},1_{\textbf{p}}}
		=
		\de{\frac{1}{\de{2\pi}^3 \sqrt{2kp}\delta(0)}}^{2}
		\De{
			1 + \coss{\de{\textbf{k}-\textbf{p}} \de{\textbf{x} -\textbf{y} } }{}
		}
		}
	which, aside from the normalization, is just the standard result of the Bose-Einstein correlation of two measurements. Now, in the classical limit, $n\approx m \gg 1$, one can show:
	\splt{
		&\bra{n_{\textbf{k}},m_{\textbf{p}}} 
		\varphi^\dagger(\textbf{y}) \varphi^\dagger(\textbf{x})
		\varphi(\textbf{x}) \varphi(\textbf{y})
		\ket{n_\textbf{k},m_{\textbf{p}}}
		\\
		&\approx
		\de{\frac{n}{\de{2\pi}^3 k \delta(0)}}^{2}  
		\De{
			1 +\frac{1}{2} \coss{\de{\textbf{k}-\textbf{p}} \de{\textbf{x} -\textbf{y} } }{}
			},
		}
	where we have assumed for simplicity that $k=p$.	
	Then, we recover the result for two localized classical sources. We see here there is still a correlation due to the Bose-Einstein symmetry, even after the classical limit.   Perhaps this is a reflection of the fact that classicalization is in realty a {\em semi-classicalization} (the effective $\hbar \rightarrow 0$ where however multiple extrema of the action retain superposition) and, as explained in the appendix, HBT is a semi-classical field theory effect.

        As a final point, we haven't found any reason to believe the correlations calculated in the last section are a result of Bose-Einstein symmetry in the same way they were for the chaotic source. 
	Therefore, our conclusion is the previous analysis in section \ref{secansatz} is unrelated to the ones made in quantum optics, meaning we aren't sure any of the techniques employed there are in fact different from the semiclassical stochastic fields examined in \cite{disorder1,disorder2} (the fact that both approaches yield the comparable effects in the trispectrum might indicate they are in fact identical).

        In conclusion, we have examined the relationship of the optical HBT effect to the dynamics of inflation.  We have found that while there are similarities between inflation and situations where HBT has been proposed, and indeed used, the differences between the two situations seem to preclude any use for the technique.  In particular, the classicalization of the quantum perturbations already at the pre-reheating stage and the off-shellness of the perturbations seem to mean that a ''chaotic'' ansatz derived here might be of limited use without a detailed assessment of how $\alpha$ in \eqref{hbtdef} varies in the slow roll phase as well as the reheating phase.
        However, we hope that the attempt illustrated here will stimulate cosmologists to explore this topic further.

      {\bf Acknowledgements}  G.T.~acknowledges support from Bolsa de produtividade CNPQ 305731/2023-8, Bolsa de pesquisa FAPESP 2023/06278-2.  GMG thanks the CAPES graduate scholarship 88887.481128/2020-00.  We thank Mike Lisa for many comments, discussions and explanations
        \appendix
\section{A coincise review of the formalism and interpretation of HBT interferometry}
        \subsection{Classical amplitude interferometry}
	
	Despite the classical description we are about to give, a more detailed analysis of the measurement procedure of the field points to it's quantum nature\cite{Glauber1963,Glauber1963a,Glauber2007,Sinyukov2013}. What one finds is that the detector doesn't react to the actual field, but rather to a complex field given by the positive frequency part of the field. So, we must expand the field through a Fourier transform and identify it's positive frequency component. Suppose a Fourier transform provides the following separation of the field
	\splt{
		\phi = \varphi + \varphi*,
		}
	such that the first term describes the positive frequency part of the field, that is contains the Fourier modes which vary with time as $e^{-i\omega t}$ with $\omega$ being the energy of the mode. The complex conjugate of $\varphi$ is called the negative frequency component of the field. Either can be used as the amplitude or complex signal for the field. As we will argue later, a detector for the field reacts to the product
	\splt{
	 \varphi^* \varphi = \abs{\varphi}^2 = I,
	} 
	which we will just call the field intensity from now on. Given a source, the field at position $x$, $\varphi(x)$, is the superposition of the sub-fields generated by each sub-source. Denoting by $\varphi_i$ the contribution from the i'th sub-field, then the full field can be written as
	\splt{
		\varphi = \sum\limits_{i} \varphi_i.
	}
	The intensity measure at some detector at position $x$ will be just
	\splt{
		I(x) = \abs{ \sum\limits_{i} \varphi_i(x) }^2.
	}
	Of course that's not the whole story, in general sources produce fields subject to some statistical uncertainty. This means what one detector sees is the averaged intensity:
	\splt{
		\mean{I(x)} 
		&= 
		\mean{\abs{ \sum\limits_{i} \varphi_i(x) }^2}.
		}
	In practice this is usually an average over time, to simplify calculations and the formalism we will work throughout this text assuming the ergodic hypothesis. So, all averages will be over some ensemble that should represent the corresponding time average. To see the interference phenomenon, we expand the averaged intensity as 	
	\splt{
		\mean{I(x)} 
	=
		\sum_i \mean{\abs{\varphi_i(x)}^2} + \sum_{i\not = j} \mean{\varphi_j^*(x) \varphi_i(x) },
		\label{AvAmp}
	}
	where $\mean{\cdots}$ denotes the ensemble average of the source. The first term in (\ref{AvAmp}) is just the sum of the averaged intensity due to each of the sub-sources, while the second term contains the interference between each pair of sub-sources at the detector as a result of their superposition at the location of the detector. So, this last term contains the information of the phase difference between each pair of sub-sources.				
	Assuming the source is chaotic, that is its sub-sources are uncorrelated and random. We implement that by supposing each sub-source has an independent random phase: 
	\splt{
		\varphi_i = \abs{\varphi_i} e^{i \alpha_i}e^{i\theta_i}.
	}
	where $\theta_i$ is a random phase. This means
	\splt{
		\mean{\varphi_i(x)} = \mean{\varphi_i(x) \varphi_j(y)} = 0
	}
	and	
	\splt{
		\mean{\varphi_i(x) \varphi^*_j(y)} 
		&=
		\mean{ 
			\abs{\varphi_i(x)} e^{i \alpha_j(x)}e^{i\theta_i}\abs{\varphi_j(y)} e^{-i \alpha_j(y)}e^{-i\theta_j}
		}
		\\
		&=
		\abs{\varphi_i(x)} e^{i \alpha_j(x)}\abs{\varphi_j(y)} e^{-i \alpha_j(y)}\mean{e^{i\de{\theta_i - \theta_j}}}
		\\
		&=
		\abs{\varphi_i(x)} \abs{\varphi_j(y)} e^{i\de{\alpha_i(x) - \alpha_j(y)}}\delta_{ij}, 
	}
	where, to obtain the above expression, we used the result for random phases
	\splt{
		\mean{e^{i \de{\theta_i - \theta_j}}} 
		&= 
		\int\limits_{0}^{2\pi}\frac{d\theta_i}{2\pi} \int\limits_{0}^{2\pi} \frac{d\theta_j}{2\pi} e^{i\de{\theta_i - \theta_j}} 
		=
		\delta_{ij}.
	}			
	Then, the intensity measured at the detector after the average will be
	\splt{
		\mean{I(x)} 
		= 
		\mean{\abs{ \sum\limits_{i} \varphi_i(x) }^2}		
		=
		\sum_{i,j} \mean{\varphi_i(x)\varphi^*_j(x)}		
		=
		\sum_{i,j} \abs{\varphi_i(x)} \abs{\varphi_j(x)} e^{i\de{\alpha_i(x) - \alpha_j(x)}} \mean{e^{i\de{\theta_i - \theta_j}}}. 
	}	
	Since the phases are random, we get
	\splt{
		\mean{I(x)}	
		=
		\sum_{i,j} \abs{\varphi_i(x)} \abs{\varphi_j(x)} \delta_{ij}
		= 
		\sum\limits_{i}\abs{\varphi_i(x)}^2 
		= 
		\sum_i I_i(x),
	}
	where $I_i$ is the intensity due to one of the sub-sources. All the relative phase information between pairs of sub-sources has been lost, which is usually where the spacetime information of the source is extracted from.
	\subsection{Classical intensity interferometry}
	The discussion on the previous section justifies why simple field measurements might present some difficulties for extracting spacetime information from a chaotic distribution of sources, since detectors will only resolve the intensities of these sources, losing the relative phase information between them.
	Fortunately, the phase information can be recovered, even for completely chaotic sources, by considering the correlation between the detection events at two different spacetime points or, equivalently, the correlation between the intensities of the field in two spacetime points:
	\splt{
		\mean{I\de{x_1} I\de{x_2}} 
		&= 
		\mean{\abs{\sum\limits_{i} \varphi_i(x_1)}^2\abs{\sum\limits_{j} \varphi_j(x_2)}^2}
		\\
		&=
		\sum_{i,j,m,n} \abs{\varphi_i(x_1)}\abs{\varphi_j(x_1)}\abs{\varphi_m(x_2)}\abs{\varphi_n(x_2)}
		\times 
		e^{i \de{\alpha_i(x_1) - \alpha_j(x_1) + \alpha_m(x_2) - \alpha_n(x_2)}}
		\mean{e^{i\de{\theta_i - \theta_j + \theta_m - \theta_n}}}.
		}
	Again, because the phases are random
	\splt{
		\mean{e^{i\de{\theta_i - \theta_j + \theta_m - \theta_n}}} = \delta_{ij}\delta_{mn} + \delta_{in} \delta_{jm} - \delta_{ij}\delta_{mn}\delta_{im},
		}	
	where the last is present to prevent double counting of the situation when $i=j=m=n$.	
	Using the above result for the phase average, we get
	\splt{
		\mean{I\de{x_1} I\de{x_2}} 
		&= 
		\sum_{i} \abs{\varphi_i(x_1)}^2 \sum_{j}\abs{\varphi_j(x_2)}^2
		\\
		&+
		\sum_{i\not=j} \abs{\varphi_i(x_1)}\abs{\varphi_j(x_1)}\abs{\varphi_i(x_2)}\abs{\varphi_j(x_2)}
		e^{i\de{\alpha_i(x_1) - \alpha_i(x_2) }}e^{-i\de{\alpha_j(x_1) - \alpha_j(x_2) }}.
		}
	Which can be simplified to
	\splt{
		\mean{I\de{x_1} I\de{x_2}} 
		&= 
		\mean{I\de{x_1}} \mean{I\de{x_2}} - \sum_{i} \abs{\varphi_i\de{x_1}}^2 \abs{\varphi_i\de{x_2}}^2  + \abs{\mean{\varphi\de{x_1}\varphi^*\de{x_2}}}^2.
		}
	Appealing to the central limit theorem\cite{Billingsley2012}, one can neglect the term $\sum_{i} \abs{\varphi_i\de{x_1}}^2 \abs{\varphi_i\de{x_2}}^2$ for a very large number of sub-sources\cite{Sinyukov2013,Aspect2020}. The resulting intensity correlation is, then, just a Gaussian random process:
	\splt{
		\mean{I\de{x_1} I\de{x_2}}
		&=
		\mean{I\de{x_1}} \mean{I\de{x_2}} + \abs{\mean{\varphi\de{x_1}\varphi^*\de{x_2}}}^2.
		}
	The last term contains the relative phase information between sub-sources. From the above, to know the correlation between the two detections, that is the measured intensity correlation independent of the individual measurements, we divide by the measured intensities in each detector:
	\splt{
		C(x_1,x_2) 
		= 
		\frac{\mean{I\de{x_1} I\de{x_2}}}{\mean{I\de{x_1}} \mean{I\de{x_2}} } 		
		=
		1
		+
		\frac{
			\abs{\mean{\varphi\de{x_1}\varphi^*\de{x_2}}}^2  
			}
		{
			\mean{I\de{x_1}}\mean{I\de{x_2}}
			}.
		}
	The last term, then, contains the correlation between the two detectors. A correlation between measurements just means that the measurement at one detector is not independent of the other one. In the classical sense, it can be seen as a consequence of the superposition principle for the fields. We will come back to this later. This correlation between intensities or detections is usually called the Hanbury Brown-Twiss(HBT) effect. If the source was coherent, that is its sub-sources didn't have random phases, then there wouldn't have been a correlation: $C(x_1,x_2) = 1$ or $\mean{I\de{x_1}I\de{x_2}} = \mean{I\de{x_1}}\mean{I\de{x_2}} $; which would just mean the two measurements are completely independent in this case. 
	
	Given very localized sub-sources, the interference seen in amplitude interferometry is not the same as an interference between the field at the locations of the two sub-sources. The detection happens at a single point and not at the location of the two sub-sources, the interference in this situation is a result of the superposition of the fields produced by each pair of sub-sources at the point of detection. So, what about intensity correlations? Where do they come from? To get a better grasp of that, consider the following rewrite of the correlation function 
	\splt{
		\mean{I\de{x_1} I\de{x_2}} 
		&= 
		\mean{\abs{\sum\limits_{i} \varphi_i(x_1)}^2\abs{\sum\limits_{j} \varphi_j(x_2)}^2}
		\\
		&=
		\mean{
			\abs{\sum_{ij} \frac{1}{2} \De{\varphi_i(x_1) \varphi_j(x_2) + \varphi_i(x_2) \varphi_j(x_1)}}^2
		}.
	}
	For chaotic sources, this results in 
	\splt{
		\mean{I\de{x_1} I\de{x_2}}
		=
		\sum\limits_{i\not = j} \abs{\frac{1}{\sqrt{2}} \De{ \varphi_i(x_1) \varphi_j(x_2) + \varphi_j(x_1)\varphi_i(x_2)  } }^2.
	}
	The above correlation is a result of the superposition of joint fields $\varphi_i\de{x_1}\varphi_j\de{x_2}$ and $\varphi_j\de{x_1}\varphi_i\de{x_2}$. The first term in this superposition corresponds to the situation in which the field at $x_1$ is excited by the i'th sub-source, while at $x_2$ it's the j'th sub-source that does the job. The second term in this superposition corresponds to the different yet indistinguishable situation where the opposite happens, $x_1$ excited by the j'th sub-source and $x_2$ by the i'th source. So, the interference phenomenon is concealed in the non-local joint measurements at $x_1$ and $x_2$ of a pair of sub-sources: $\abs{\varphi_i\de{x_1}\varphi_j\de{x_2} + \varphi_j\de{x_1}\varphi_i\de{x_2}}^2$. It's this superposition that survives the destructive interference originating from the chaotic nature of the source, implying a non-local interference of the field. This raises some physical questions: How can two independent detections influence one another in a classical theory? Shouldn't the superpositions at the two detectors be independent? This interference seems to be beyond classical physics because of the apparent non-locality of the effect\cite{Shih2020,Scarcelli2006}. Although a bit controversial, a good explanation can be provided in the far-field limit\cite{Silverman2008} but not for near-field\cite{Shih2020,Scarcelli2006}. Perhaps the best explanation for the HBT effect relies on quantum mechanics, as we are about to see.

	On the other hand, in the continuous limit
	\splt{
		\mean{I\de{x_1} I\de{x_2}}
		=
		\int d^4y_1 d^4y_2 \abs{ \frac{1}{\sqrt{2}} \De{ \varphi(y_1,x_1)\varphi(y_2,x_2) + \varphi(y_2,x_1)\varphi(y_1,x_2)  }  }^2.
		}  	
	

	Under the assumption that $\abs{\varphi_i(x)} = \abs{\varphi_i}$, we can write it as
	\splt{
		\mean{I\de{x_1} I\de{x_2}}
		&=
		\frac{1}{2}
		\mean{
			\abs{
				\sum\limits_{ij} \abs{\varphi_i} \abs{\varphi_j} e^{i\de{\theta_i + \theta_j}} 
				\psi_{i,j}(x_1,x_2)
				}^2
			}
		}
	This can be simplified to
	\splt{
		\mean{I\de{x_1} I\de{x_2}}
		&=
		\sum\limits_{i\not = j} \abs{\varphi_i}^2 \abs{\varphi_j}^2 \abs{\psi_{i,j}(x_1,x_2)}^2 + \sum\limits_i \de{\abs{\varphi_i}^2}^2.
		}		
	Which is the same result as before.	For the continuous limit, on the other hand, we can rewrite the correlation between the intensities as
	\splt{
		\mean{I\de{x_1} I\de{x_2}} = \int d^4 y_1 d^4 y_2 \abs{\varphi\de{y_1}}^2 \abs{\varphi\de{y_2}}^2 \abs{ \psi\de{y_1,y_2,x_1,x_2} }^2,
		}
	where
	\splt{
		\psi\de{y_1,y_2, x_1,x_2}
		\frac{1}{\sqrt{2}}
		\De{
			e^{i\alpha\de{y_1,x_1}}e^{i\alpha\de{y_2,x_2}} 
			+
			e^{i\alpha\de{y_2,x_1}}e^{i\alpha\de{y_1,x_2}}
			}.
		}	
	These expressions will come in handy when we talk about the quantum interpretation.

	\subsection{Quantum theory of interferometry}
	
	A detector is a quantum system that interacts with the field and performs a transition, usually it ejects an electron with some probability into some circuitry. When this interaction is small enough one can use perturbation theory. To first order in perturbations, one can show that the rate of transition amplitude is given by\cite{Glauber2007,Klauder2006,Mandel1995}
$		D(x)\bra{f} \varphi\de{x}\ket{i}$.
		
	for an ideal detection localized at spacetime point $x$. On this expression $\ket{i}$ is the state of the field before detection, $\ket{f}$ is the state of the field after detection and $D$ is a complex number containing all the information from the detector transition. The only term from the field present is its annihilation part, since the detector is assumed to be in its ground state and so the absorption term is dominant. Now, all we measure is the total counting rate on the detector and not the final state of the field. So, we must trace out the final state of the field to calculate the probability for the rate of transition\cite{Glauber2007,Klauder2006,Mandel1995}:
	\splt{
		P_1(x) 
		= 
		\abs{D(x)}^2 \sum_f \bra{i} \varphi^\dagger\de{x} \ket{f}\bra{f} \varphi\de{x} \ket{i}
		=
		\abs{D(x)}^2 \bra{i} \varphi^\dagger\de{x}\varphi\de{x} \ket{i}.
		}
	The counting rate $P_1(x)$ is simply the probability per unit time that the ideal detector in question absorbs a particle at the spacetime point $x$. Also, the initial state doesn't need to be pure. For a generic state, we suppose the field is in the state described by some density matrix $\rho_i$. Then, the counting rate becomes
	\splt{
		P_1(x) = \abs{D\de{x}}^2 \text{Tr}\de{\rho_i \varphi^\dagger\de{x} \varphi\de{x}},
		}
	where $\text{Tr}$ denotes the trace over the state space. The above is the kind of observable amplitude interferometers use, since they involve a single detector, measuring the intensity of the field $\mean{I}=\mean{\varphi^\dagger \varphi}$.
	
	Intensity or HBT interferometers contain two detectors. In this situation, then we stop at second order in perturbation theory to get, for two ideal detection events at $x$ and $y$\cite{Glauber2007,Klauder2006,Mandel1995} is
	$D(y)D(x) \bra{f} \varphi(y) \varphi(x)\ket{i}.$
	
	One could then assume the above is time ordered, $x^0<y^0$. Following the same procedure as before, we find the probability of joint transition rate to be\cite{Glauber2007,Klauder2006,Mandel1995}
	\splt{
		P_2(x,y) = \abs{D\de{x}}^2 \abs{D\de{y}}^2 \text{Tr}\de{\rho_i \varphi^\dagger\de{x}\varphi^\dagger\de{y}\varphi\de{y}\varphi\de{x} }.
		\label{Int_Cor}
		}
	This is	the observable used in intensity or HBT interferometry. From (\ref{Int_Cor}), we can already get a good grasp on how quantum mechanics naturally supports correlations between two detections.

        Consider the effect the measurement of a particle at spacetime point $x$ has in the state of the field:
	$	\varphi\de{x} \rho_i \varphi^\dagger(x).$:
	The resulting state after the measurement is given by the reduced density matrix\cite{Agarwal2013}
	\splt{
		\rho_{i,\text{red}}(x) = \frac{\varphi\de{x} \rho_i \varphi^\dagger(x)}{\text{Tr}\de{\rho_i \varphi^\dagger(x) \varphi(x)}}.
		}
	So, if $x^0<y^0$, we can interpret $\text{Tr}\de{\rho_i \varphi^\dagger(x)\varphi^\dagger(y)\varphi(y)\varphi(x)}$ as the measurement of a particle at spacetime point $y$ conditioned by the detection of a particle at $x$:
	\splt{
		\text{Tr}\de{\rho_i \varphi^\dagger(x)\varphi^\dagger(y)\varphi(y)\varphi(x)}
		&=
		\text{Tr}\de{\varphi(y) \rho_{i,\text{red}} \varphi^\dagger(y) }\text{Tr}\de{\varphi(x)\rho_i \varphi^\dagger(x)};
		}
	where in the first line we used the cyclic property of the trace. Therefore, if
	\splt{
		\rho_{i,\text{red}}(x) \not = \rho_i,
		}
	then a detection of another particle at $y$ will be influenced by the first measurement at $x$, since the first detection has changed the state of the field. As we will see, this is the reason for the correlations encountered when performing a HBT type of measurement with a chaotic source. Of course one could also stack up more measurements and build whatever order of correlation they want\cite{Glauber1963}. For our purposes, since we only want to talk about HBT kind of measurements, it's enough to stop at second order.
	
	\subsection{Coherent states and quantum fields interacting with a classical source}
	
	A very important kind of state for quantum optics and interferometry are coherent states $\ket{\alpha}$. They are defined by the property of being eigenstates of the annihilation operator\cite{Glauber1963,Ballentine2014,Mandel1995,Klauder2006}:
	\splt{
		a\de{\textbf{k}}\ket{\alpha} = \alpha\de{\textbf{k}} \ket{\alpha};
		\label{CoherentState}
		}
	where the eigenvalue $\alpha\de{\textbf{k}}$ is a complex number. We can solve (\ref{CoherentState}) by assuming it can be created from the vacuum and treating formally the action of an annihilation operator as a derivative with respect to the creation operator. That is, we rewrite (\ref{CoherentState}) as
	\splt{
		\frac{\delta }{\delta a^{\dagger}\de{\textbf{k}}} D\de{\alpha, a^\dagger} \ket{0} = \alpha\de{\textbf{k}} D\de{\alpha,a^\dagger }\ket{0}.
		}
	Which has the solution
	\splt{
		D\de{\alpha, a^\dagger} = e^{-\frac{\de{2\pi}^3}{2} \int d^3k \abs{\alpha\de{\textbf{k}}}^2 } e^{\int d^3 k \alpha\de{\textbf{k}} a^{\dagger}\de{\textbf{k}} }\ket{0},
		}
	where the normalization was obtained by assuming 
	\splt{
		\inner{\alpha}{\alpha} = 1
		}
	and using the Baker-Hausdorff-Campbell formula\cite{Hall2003}. Coherent states are extremely important for a variety of reasons. One of them is that they form an over-complete set of states\cite{Mandel1995,Scully1997,Glauber2007}
	\splt{
		\int \mathcal{D}\alpha \ket{\alpha}\bra{\alpha}, 
		\quad 
		\inner{\beta}{\alpha} 
		= 
		e^{-\frac{1}{2} \int d^3k \de{\abs{\alpha\de{\textbf{k}}}^2 + \abs{\beta\de{\textbf{k}}}^2  + \beta^*\de{\textbf{k}}\alpha\de{\textbf{k}}  }},
		}
	where
	\splt{
		\mathcal{D}\alpha = \prod\limits_{\textbf{k} \in \mathbb{R}^3}\frac{ d^2 \alpha\de{\textbf{k}}}{\pi}.
		}
	They are also the states that best describe classical fields\cite{Glauber2007,Mandel1995,Ficek2016,Scully1997}, with the classical field identified as
	\splt{
		\phi_{\alpha}(x) 
		&= 
		\bra{\alpha}\phi(x)\ket{\alpha}
		\\
		&=
		\int \frac{d^3k}{\de{2\pi}^3} \frac{e^{ik\cdot x}}{\sqrt{2\omega\de{k}}} \alpha\de{\textbf{k}} + \int \frac{d^3k}{\de{2\pi}^3} \frac{e^{-ik\cdot x}}{\sqrt{2\omega\de{k}}}\alpha^*\de{\textbf{k}}.
		\label{ClasField}
		} 
	We see the above gets the same shape of a classical field if we identify its complex signal or amplitude with $\varphi_\alpha$. More importantly than that, they can be used to define what are classical and non-classical states. By a classical state, we mean a state with a direct classical field analog, such as (\ref{ClasField}). To establish this criterion, one first need the fact that any state can be expanded as\cite{Sudarshan1963,Mandel1995,Mandel1986,Mehta1967,Glauber1963a}
	\splt{
		\rho = \int \mathcal{D}\alpha \mathcal{P}\de{\alpha} \ket{\alpha}\bra{\alpha}.
		}
	Called the Glauber-Sudarshan representation or just the P-representation of $\rho$. From this representation, one can say a state $\rho$ is classical if $\mathcal{P}\de{\alpha}$ is a classical probability distribution, meaning it's non-negative\cite{Mandel1986,Paul1982}. Otherwise the state is said to be non-classical, since there is no classical analog to the observables of this quantum field. From the above discussion, it follows we must be able to represent the sources in the classical context, described in the previous section, as a classical density matrix. We will see that is very natural in the following.	
	
	Interestingly there is a very simple way to generate coherent states. The state produced from the vacuum by the interaction of a classical source with a quantum field is a coherent state. This interaction is described by the Lagrangian\cite{Nastase2019,Das2020,Chen2018}
	\splt{
		\mathcal{L} = -\frac{1}{2}\partial \phi(x) \cdot \partial \phi(x) + J(x) \phi(x),
	}
	where $J$ is the classical source assumed to be a well behaved spacetime function which goes to zero in spacelike and timelike directions. With this Lagrangian, the field satisfies 
	\splt{
		\partial^2 \phi = - J.
	}
	The solution, under causal boundary conditions, should be\footnote{We will sometimes use $A\cdot B$ to represent a volume integral over the space $A$ and $B$ are functions of. An example is $\phi \cdot J = \int d^4x \phi(x)J(x)$.}
	\splt{
		\phi(x) = \phi_0 + i D_R\cdot J(x) = \phi_0 + i \int d^4y D_R(x-y)J(y),
		}
	where $\phi_0$ is the free solution to the Klein-Gordon equation and $D_R$ is the retarded propagator, i.e. the inverse of $\partial^2$:
	\splt{
		\partial^2(x) D_R(x-y) = i \delta(x - y);
		}
	with causal boundary conditions
	\splt{
		D_R(x-y) = 0, \text{ when } x^0<y^0.
		}
	As a function it can be written
	\splt{\label{wightdef}
		D_R(x-y) = \Theta(x^0 - y^0) \de{K(x-y) - K(y-x) },
		}
	where $K$ is the kernel or Wightman function of $\partial^2$, that is the on-shell function
	\splt{
		K(x - y) = \int \frac{d^3 k }{\de{2\pi}^3} \frac{e^{i k\cdot \de{x-y}}}{2k}.
		}
	If we assume the source is in the past of the point we want to evaluate the field in, $x^0 > y^0$, then the full solution of the field can be expressed as
	\splt{
		\phi(x) 
		= 
		\phi_0(x) + i K\cdot J(x) + \de{iK\cdot J(x)}^*
		=
		\phi_0(x) + \varphi_J(x) + \varphi^*_J(x).
		}					
	This is the full solution in the Heisenberg picture. 
	
	We could just identify the complex signal of classical field produced by the source with $\varphi_J$, since
	\splt{
		\bra{0} \phi(x)\ket{0} = \varphi_J(x) + \varphi^*_J(x),
		}
	and work from there in the Heisenberg picture. But, a more useful approach for interferometry is the one derived in the interaction picture\cite{Nastase2019,Schwartz2014,Das2020}. 
	Denoting the field in the interaction picture as $\phi_I$ and treating $ \mathcal{L}_I = \phi_I(x) J(x)$ as an interaction, the evolution operator in the interaction picture takes the form\cite{Nastase2019,Schwartz2014,Das2020,Chen2018}
	\splt{
		U_I
		= 
		T \De{
			\exp\de{
				i\int d^4x \mathcal{L}_I
			}
		}
		=	
		T \De{
			\exp\de{
				i\int d^4x J(x) \phi_I(x).
			}
		}	
	\label{EvoOp}	
	}
	By assuming the field starts out in the vacuum state, we calculate the evolved state using Wick's theorem. In it's most general form in the interaction picture, it states that\cite{Chen2018,Diosi2018}
	\splt{
		T(\mathcal{O}(\phi_I)) =  :\exp\de{\frac{1}{2}\int d^4 x d^4y \DD{}{\phi_I(x)} D_F(x,y) \DD{}{\phi_I(y)} } \mathcal{O}(\phi_I) :,
		\label{wick}
	}
	where $:\mathcal{O}:$ denotes normal ordering of some operator $\mathcal{O}$ and $D_F$ is the Feynman propagator. Applying (\ref{wick}) to the evolution operator (\ref{EvoOp}), one gets
	\splt{
		U_I 
		&=
		:\exp\de{\frac{1}{2}\int d^4 x d^4y \DD{}{\phi_I(x)} D_F(x,y) \DD{}{\phi_I(y)}} \exp\de{i \int d\tau d^3 x J (x) \phi_I(x)} :
		\\
		&=
		\exp\De{ -\frac{1}{2} \int d^4 x d^4 y J(x) D_F(x,y) J(y) } :\exp\de{i \int d\tau d^3 x J (x) \phi_I(x)}:.
	}
	Expanding the field in annihilation and creation operators, we have
	\splt{
		:\exp\de{i \int d^4 x J (x) \phi_I(x)}: 
		&=
		\exp\de{\int \frac{d^3 k}{\de{2\pi}^3} \tilde{J}(\textbf{k}) a^\dagger_{\textbf{k}} } \exp\de{\int \frac{d^3 k}{\de{2\pi}^3} \tilde{J}^*(\textbf{k})a_{\textbf{k}}},
	}
 Finally, the evolution of the vacuum state will be
	\splt{
		\ket{J} 
		= 
		U_I \ket{0}
		&=
		\exp\De{ -\frac{1}{2} \int d^4 x d^4 y J(x) D_F(x,y) J(y) } \exp\de{\int \frac{d^3 k}{\de{2\pi}^3} \tilde{J}(\textbf{k}) a^\dagger_{\textbf{k}} } \ket{0}.
	}
	Since $U_I$ is unitary, $\ket{\tilde{J}}$ must be normalized to one. So, it's just a coherent state up to some irrelevant phase. From the normalization and the Baker-Campbell-Hausdorff formula\cite{Hall2003}, one can show
	\splt{
		\text{Real}\de{\int d^4 x d^4 y J(x) D_F(x,y) J(y)} =  \int \frac{d^3k}{\de{2\pi}^3} \abs{\tilde{J}(\textbf{k})}^2,
	}
	that isn't hard to do\cite{Chen2018}. 
	Therefore, up to an irrelevant phase, the coherent state produced by a classical source can be written as
	\splt{
		\ket{\tilde{J}} = \exp\de{ -\frac{1}{2}  \int \frac{d^3 k}{\de{2\pi}^3} \abs{\tilde{J}(\textbf{k})}^2 } \exp\de{\int \frac{d^3 k}{\de{2\pi}^3} \tilde{J}(\textbf{k}) a^\dagger_{\textbf{k}} } \ket{0},
	}
	where $\tilde{J}\de{\textbf{k}}$ is the on-shell Fourier transform of the source.	
	\subsection{Quantum state and density matrix of Chaotic sources}
	The same construction of the classical theory can be made for a source in the quantum theory. Let's assume our source can be split into many sub-sources
\splt{J(x) = \sum\limits_{i} J_i(x).}
To make the source chaotic, just add a random phase to each $\tilde{J}_i$:
	\splt{
		\tilde{J}\de{\textbf{k}} 
		= 
		\sum\limits_{i} \tilde{J}_i\de{\textbf{k}} e^{i\theta_i}.
		}
	As we saw before, a very large number of sources just makes this system into a Gaussian random process. These are completely determined by their two point function:
	\splt{
		\mean{\tilde{J}^*(\textbf{k})\tilde{J}(\textbf{p})} 
		= 
		\sum\limits_{i} \tilde{J}^*_i\de{\textbf{k}}\tilde{J}_i\de{\textbf{p}}.
	}
     
	So, our chaotic source can be described by the following density matrix
	\splt{
		\rho_{\text{ch}} = \int \mathcal{D}\tilde{J} \exp\de{-\tilde{J}^* \cdot M^{-1}\cdot \tilde{J}} \ket{\tilde{J}} \bra{\tilde{J}},
		}
	where	
	\splt{
		M(\textbf{k},\textbf{p}) 
		= 
		\mean{\tilde{J}^*(\textbf{k})\tilde{J}(\textbf{p})} 	
		=
		\sum\limits_{i} \tilde{J}^*_i\de{\textbf{k}} \tilde{J}_i\de{\textbf{p}}.
		}
	To check that, just consider the generating functional for normally ordered products of annihilation and creation operators
	\splt{
		Z\de{f,f^*} 
		=
		\text{Tr} \de{ \rho_{\text{ch}} e^{f\cdot \varphi^\dagger}e^{f^*\cdot \varphi} } 
		=
		\int \mathcal{D}\tilde{J}  \exp\de{-\tilde{J}^* \cdot M^{-1}\cdot \tilde{J}} \exp\de{\tilde{f}^*\cdot \tilde{J} + \tilde{f} \cdot \tilde{J}^*},
		\label{GenFunc}
		}
	where $\tilde{f}$ is again just the on-shell Fourier transform of the auxiliary function $f$. One can solve the Gaussian integral (\ref{GenFunc}) exactly as
	\splt{
		Z\de{f,f^*} = \exp\de{ \tilde{f}\cdot M \cdot \tilde{f}^*},
		}
	where we have assumed that $Z(0,0) = 1$. Hence, any average of products of $\tilde{J}^*$ and $\tilde{J}$ can be determined by derivatives with respect to $\tilde{f}$ and $\tilde{f}^*$: 
	\splt{
		\text{Tr}\de{\rho_{\text{ch}} \prod_i a^\dagger\de{\textbf{k}_i} \prod_j a\de{\textbf{p}_j}}
		&=
		\mean{\prod_i \tilde{J}^*\de{\textbf{k}_i} \prod_j \tilde{J}\de{\textbf{p}_j}}
		}
		\[\
		\int \mathcal{D}\tilde{J}  \exp\de{-\tilde{J}^* \cdot M^{-1}\cdot \tilde{J}}
		\prod_i \tilde{J}^*\de{\textbf{k}_i} \prod_j \tilde{J}\de{\textbf{p}_j}
		= 
		\left.\prod_i \frac{\delta}{\delta\tilde{f}\de{\textbf{k}_i}}
		\prod_j \frac{\delta}{\delta\tilde{f}^*\de{\textbf{k}_j}}
		Z(f,f^*)\right\vert_{f,f^*=0}. \]
	in configuration space, using the Fourier transform properties of the Gaussian
	\splt{
		\text{Tr}\de{\rho_{\text{ch}} \prod_i \varphi^\dagger\de{x_i} \prod_j \varphi\de{y_j}}
		&=
		\mean{ \prod_i \varphi_J^*\de{x_i} \prod_j \varphi_J\de{y_j} }
		}\[\
		=
		\int \mathcal{D}\tilde{J}  \exp\de{-\tilde{J}^* \cdot M^{-1}\cdot \tilde{J}}
		\prod_i \varphi_J^*\de{x_i} \prod_j \varphi_J\de{y_j}
		= 
		\left.\prod_i \frac{\delta}{\delta f\de{x_i}}
		\prod_j \frac{\delta}{\delta f^*\de{y_j}}
		Z(f,f^*)\right\vert_{f,f^*=0}.
		\]
	These relations make it easy to calculate all the correlation functions. For example, the correlation function for two particle detections can be readily calculated
	\splt{
		\text{Tr}\de{\rho_{\text{ch}} \varphi^\dagger(x)\varphi^\dagger(y)\varphi(y)\varphi(x)}
		&=
		\mean{\abs{\varphi_J(x)}^2}\mean{\abs{\varphi_J(y)}^2} + \abs{\mean{\varphi_J(x) \varphi^*_J(y) }}^2
		\\
		&=
		\text{Tr}\de{\rho_{\text{ch}}\varphi^\dagger(x)\varphi(x)}\text{Tr}\de{\rho_{\text{ch}}\varphi^\dagger(y)\varphi(y)} + \abs{\text{Tr}\de{\rho_{\text{ch}} \varphi^\dagger(y)\varphi(x)}}^2.
		}
	Which is just the same result obtained in the classical theory.	
	
	A few points are in order. If the source wasn't chaotic, then it would simply be a coherent state: $\rho_{\text{co}} = \ket{\tilde{J}} \bra{\tilde{J}}$. This implies that the correlation functions are all factorisable
	\splt{
		\text{Tr}\de{\rho_{\text{co}} :\prod_i \varphi^\dagger\de{x_i} \varphi\de{x_i}:}
	=
		\prod_i \abs{\varphi_J\de{x_i}}^2
	        =
		\prod_i\text{Tr}\de{\rho_{\text{co}} \varphi^\dagger\de{x_i} \varphi\de{x_i}}
		,
		}
	simply because of the fact that coherent states are eigenstates of the annihilation operators. Immediately we know all the normalized correlation functions $C(x_1,\cdots,x_n)$ are equal to one. For coherent states, therefore, all the measurements made are completely independent from the others. That can be easily seen by considering the effect of a measurement of a field observable $\varphi(x)$ on the state
	\splt{
		\rho_{\text{co},\text{red}}(x) 
		&=
		\frac{\varphi(x) \rho_{\text{co}} \varphi^\dagger(x)}{\text{Tr} \de{\rho \varphi^\dagger(x)\varphi(x) } }
	= 	\frac{\varphi(x) \ket{\tilde{J}} \bra{\tilde{J}} \varphi^\dagger(x) }{\text{Tr} \de{\ket{\tilde{J}} \bra{\tilde{J}} \varphi^\dagger(x)\varphi(x) } }.	
	=	\ket{\tilde{J}} \bra{\tilde{J}},
	}	 	
	where we have used in the last line that trace is linear and that coherent states are normalized to one: $\text{Tr}\de{\ket{\tilde{J}} \bra{\tilde{J}} } = 1$. As expected, a measurement didn't change the density matrix, 
	proving the previous statement on coherent states.  

        The same won't be true for a chaotic source. The density matrix changes after a measurement
	\splt{
		\rho_{\text{red},\text{ch}}(x) 
		=
		\frac{1}{\mean{\abs{\varphi_J(x)}^2 } } \int \mathcal{D}\tilde{J} \exp\de{-\tilde{J}^* \cdot M^{-1}\cdot \tilde{J}}\abs{\varphi_J(x)}^2  \ket{\tilde{J}} \bra{\tilde{J}}
		\ne \rho_{\text{ch}}.
	}
	That can also be seen from the fact that the correlation function doesn't factorize
	\splt{
		\text{Tr}\de{\rho_{\text{ch}} \varphi^\dagger(x)\varphi^\dagger(y)\varphi(y)\varphi(x)}
		\not =
		\text{Tr}\de{\rho_{\text{ch}} \varphi^\dagger(x)\varphi(x)}
		\text{Tr}\de{\rho_{\text{ch}}\varphi^\dagger(y)\varphi(y)}.
	}
        While in a sense this is a trivial consequence of the coherent state being an Eigenstate of the field operator and the chaotic state not being one, 
	these results may still seem counterintuitive: how can a coherent source with no fluctuations in their amplitudes lead to uncorrelated measurements? Coherent states, being the least random states of the field, emit particles at regular intervals. A chaotic source, on the other hand, emits more particles when its amplitude is high and fewer when low. This bunching of particles as the amplitude is at a high point is the reason for the correlations observed for a chaotic source. Therefore, statistical dependence is a characteristic of incoherence in this context while coherence results in statistical independence.

        This also clarifies the relation of interferometry to statistical mechanics, of which random phases are an expected feature \cite{berry}.
        In Global equilibrium, the density matrix is ''chaotic'' with random phases, as expected from maximal mixing.   However, in the thermodynamic limit the amplitude for emitting particles is uniform across the (large) volume of the system and constant in time because of the KMS state\cite{kms}.  It is therefore understandable that there is no interferometry, as clear from the fact that in the Gaussian approximation \cite{heinz,lisa} the HBT radii and emission times are infinite, hence correlation radii are zero.   The gradient in bunchings reflects the fact that {\em local} equilibrium is not full equilibrium, since gradients will relax in the long run.

        HBT-type correlations are however more general than chaotic, coherent or thermal sources. As long as the correlation function does or doesn't factorize the system won't or will have the HBT effect, in the sense of detections not being independent.  
Finally, in the final subsection we will go into more detail on the quantum mechanical reason for the HBT effect.

	\subsection{Quantum interpretation of the HBT effect}
		
	%
	
	The original HBT experiment\cite{Brown1954,BROWN1956,Brown1974,Silverman2008} worked by correlating the fluctuations of the currents from two photon detectors illuminated by a chaotic source of light. If one represents each current by\cite{Silverman2008}
	\splt{
		i_j(t) = \bar{i}_j + \Delta i_j(t)
		}
	where $\bar{i}_j$ is the stationary average of the current and $\Delta i_j$ is the fluctuation, then the correlation between the two currents can be expressed as
	\splt{
		C 
		= 
		\frac{\mean{i_1(t_1) i_2(t_2) }}{\bar{i}_1 \bar{i}_2}
		=
		1 + \frac{\mean{\Delta i_1(t_1) \Delta i_2(t_2) }}{\bar{i}_1 \bar{i}_2}.
		}
	The brackets $\mean{\cdots}$ represents an average over time. Their result was simply that 
	\splt{
		\frac{\mean{\Delta i_1(t_1) \Delta i_2(t_2) }}{\bar{i}_1 \bar{i}_2}>0,
		}
	demonstrating the two signals were correlated. This result surprised the physics community at the time. The original version of the experiment wasn't necessarily controversial, it can be explained through classical electromagnetism\cite{Shih2020,Scarcelli2006,Silverman2008}, but rather a variation of it. If the source is sufficiently weak one could count the detected particles and look for correlations in their arrivals. The HBT result seemed to imply there would be correlations on this version of the experiment as well. The controversy comes from the implication that correlated arrivals of two different photons, emitted randomly from the source, was impossible since these two photons propagating freely in empty space couldn't possibly interfere with one another. From Dirac's \emph{Principles of quantum mechanics}\cite{Dirac1981}: ``Each photon interferes only with itself. Interference between two different photons never occurs''. Two different photons interfering was, as people believed, against the laws of quantum mechanics\cite{Brown1974}. As was first realized by Glauber\cite{Aspect2020,Glauber2007}, that's not the case at all, although one should talk about {\em quantum field theory} (second quantization) rather than quantum mechanics. The HBT effect follows from standard quantum formalism as will be shown below.  The ``classical'' HBT will simply be the semiclassical limit (It is sometimes said that ''there is no H in HBT''; What happens is that Gaussian emission function exponents are {\em divided} by $\hbar$ to be dimensionless, rather than multiples of $\hbar$).
	
	The first observation one must make, strictly speaking, it's not the photons that are interfering, since they are not the fundamental objects, but the electromagnetic field. What is interfering are the alternative ``histories'' of the whole system. To see how that follows from the formalism established in the last section for a scalar field, consider a source composed of only two sub-sources
	\splt{
		J(x) = J_1(x) + J_2(x),
		} 
	then the field amplitude will also have only two contributions, one for each sub-source, 
	\splt{
		\varphi(x) 
		&= 
		i \int d^4y \int \frac{d^3k}{\de{2\pi}^3} \frac{e^{i k\cdot\de{x - y}}}{2k} J_1(y)
		+
		i \int d^4y \int \frac{d^3k}{\de{2\pi}^3} \frac{e^{i k\cdot\de{x - y}}}{2k} J_2(y)
		\\
		&=
		\varphi_{J_1}(x) + \varphi_{J_2}(x),
		}
	where $\varphi_{J_i}(x)$ can be interpreted as the amplitude for a particle to be emitted by the sub-source $J_i$ and be measured at the spacetime point $x$. So, the amplitude for joint detections at $x_1$ and $x_2$ is then
	\splt{
		\bra{f} \varphi(x_2) \varphi(x_1)\ket{\tilde{J}} 
		=
		\de{ \varphi_{J_1}(x) + \varphi_{J_2}(x) } \de{ \varphi_{J_1}(x) + \varphi_{J_2}(x) } \inner{f}{\tilde{J}}.
		}
	There are four possible amplitudes:
	\begin{itemize}
		\item[(1)] Both particles got emitted by sub-source $J_1$.
		\item[(2)] Both particles got emitted by sub-source $J_2$.
		\item[(3)] Each particle was emitted by a different sub-source. Because of the bosonic nature of them, there are two indistinguishable possibilities:
		\begin{itemize}
			\item[(3.1)] The particle emitted by $J_1$ was detected at $x_1$, while the one emitted by $J_2$ got detected at $x_2$.
			\item[(3.2)] The particle emitted by $J_1$ was detected at $x_2$, while the one emitted by $J_2$ got detected at $x_1$.
		\end{itemize}
	\end{itemize}
	The Bose-Einstein symmetry responsible for the last two amplitudes arises automatically in the source formalism we are using here. These four amplitudes can be represented diagrammatically as
	\splt{
		\de{
		\begin{tikzcd} [ampersand replacement=\&]
			J_1 \ar[r, dash]\& x_1
			\\
			J_2 \ar[r, dash]\& x_2
		\end{tikzcd}
		+
		\begin{tikzcd} [ampersand replacement=\&]
			J_1 \ar[dr, dash] \& x_1
			\\
			J_2\ar[ur, dash, crossing over] \& x_2
		\end{tikzcd}
		+
		\begin{tikzcd} [ampersand replacement=\&]
			J_1 \ar[r, dash]\ar[dr, dash]\& x_1
			\\
			J_2 \& x_2
		\end{tikzcd}
		+
		\begin{tikzcd} [ampersand replacement=\&]
			J_1 \& x_1
			\\
			J_2 \ar[r, dash]\ar[ur, dash]\& x_2
		\end{tikzcd}
		}
		\inner{f}{\tilde{J}},
		}
	where we are representing
	\splt{
		\varphi_{J_i}\de{x_j} 
		= 	
		\begin{tikzcd} [ampersand replacement=\&]
			J_i \ar[r, dash]\& x_j
		\end{tikzcd}.
		}
	To calculate the probability, we trace out the final state of the field and get
	\splt{
		&\text{Tr}\de{\rho\varphi^\dagger(x)\varphi^\dagger(y)\varphi(y)\varphi(x)} 
		\\
		&=
		\mean{
		\abs{
			\begin{tikzcd} [ampersand replacement=\&]
				J_1 \ar[r, dash]\& x_1
				\\
				J_2 \ar[r, dash]\& x_2
			\end{tikzcd}
			+
			\begin{tikzcd} [ampersand replacement=\&]
				J_1 \ar[dr, dash] \& x_1
				\\
				J_2\ar[ur, dash, crossing over] \& x_2
			\end{tikzcd}
			+
			\begin{tikzcd} [ampersand replacement=\&]
				J_1 \ar[r, dash]\ar[dr, dash]\& x_1
				\\
				J_2 \& x_2
			\end{tikzcd}
			+
			\begin{tikzcd} [ampersand replacement=\&]
				J_1 \& x_1
				\\
				J_2 \ar[r, dash]\ar[ur, dash]\& x_2
			\end{tikzcd} 
		}^2
		}.
	}
	If the two sub-sources are statistically independent and random, then after the phase average
	\splt{	
		&\text{Tr}\de{\rho_{\text{ch}}\varphi^\dagger(x)\varphi^\dagger(y)
			\varphi(y)\varphi(x)} 
		\\
		&=
		\mean{
		\abs{
		\begin{tikzcd} [ampersand replacement=\&]
			J_1 \ar[r, dash]\& x_1
			\\
			J_2 \ar[r, dash]\& x_2
		\end{tikzcd}
		+
		\begin{tikzcd} [ampersand replacement=\&]
			J_1 \ar[dr, dash] \& x_1
			\\
			J_2\ar[ur, dash, crossing over] \& x_2
		\end{tikzcd}
		}^2	
		}	
		+
		\mean{
		\abs{	
		\begin{tikzcd} [ampersand replacement=\&]
			J_1 \ar[r, dash]\ar[dr, dash]\& x_1
			\\
			J_2 \& x_2
		\end{tikzcd}
		}^2
		}	
		+
		\mean{
		\abs{
		\begin{tikzcd} [ampersand replacement=\&]
			J_1 \& x_1
			\\
			J_2 \ar[r, dash]\ar[ur, dash]\& x_2
		\end{tikzcd} 
		}^2
		}.
	}
	Therefore, the only interference come from the indistinguishable amplitudes
	\splt{
		&\mean{
			\abs{
				\begin{tikzcd} [ampersand replacement=\&]
					J_1 \ar[r, dash]\& x_1
					\\
					J_2 \ar[r, dash]\& x_2
				\end{tikzcd}
				+
				\begin{tikzcd} [ampersand replacement=\&]
					J_1 \ar[dr, dash] \& x_1
					\\
					J_2\ar[ur, dash, crossing over] \& x_2
				\end{tikzcd}
			}^2	
		}
		=
		\mean{
			\abs{
				\begin{tikzcd} [ampersand replacement=\&]
					J_1 \ar[r, dash]\& x_1
					\\
					J_2 \ar[r, dash]\& x_2
				\end{tikzcd}
				}^2
			}
		+
		\mean{
			\abs{
				\begin{tikzcd} [ampersand replacement=\&]
					J_1 \ar[dr, dash] \& x_1
					\\
					J_2\ar[ur, dash, crossing over] \& x_2
				\end{tikzcd}
			}^2
		}
		\\
		&+
		\de{
			\begin{tikzcd} [ampersand replacement=\&]
				J_1 \ar[r, dash]\& x_1
				\\
				J_2 \ar[r, dash]\& x_2
			\end{tikzcd}
		}
		\de{
			\begin{tikzcd} [ampersand replacement=\&]
				J_1 \ar[dr, dash] \& x_1
				\\
				J_2\ar[ur, dash, crossing over] \& x_2
			\end{tikzcd}
			}^*
		+
		\de{
			\begin{tikzcd} [ampersand replacement=\&]
				J_1 \ar[r, dash]\& x_1
				\\
				J_2 \ar[r, dash]\& x_2
			\end{tikzcd}
		}^*
		\de{
			\begin{tikzcd} [ampersand replacement=\&]
				J_1 \ar[dr, dash] \& x_1
				\\
				J_2\ar[ur, dash, crossing over] \& x_2
			\end{tikzcd}
		},		
	}
	where the last two are the non-local interference terms\cite{Valentini1991}. The HBT effect or the correlations for chaotic sources are a result of this constructive interference which resists the phase average. That can easily be seen by expressing in the same diagrammatic way the expression for two independent measurements:
	\splt{
		&\text{Tr}\de{\rho_{\text{ch}}\varphi^\dagger(x_1)\varphi(x_1)}\text{Tr}\de{\rho_{\text{ch}}\varphi^\dagger(x_2)\varphi(x_2)}
		\\
		&=
		\mean{
			\abs{
				\begin{tikzcd} [ampersand replacement=\&]
					J_1 \ar[dr, dash] \& 
					\\
					\& x_2
				\end{tikzcd}
				+
				\begin{tikzcd} [ampersand replacement=\&]
					\\
					J_2 \ar[r, dash]\& x_2
				\end{tikzcd}
			}^2	
		}
		\mean{
			\abs{
				\begin{tikzcd} [ampersand replacement=\&]
					J_1 \ar[r, dash] \& x_1 
					\\
				\end{tikzcd}
				+
				\begin{tikzcd} [ampersand replacement=\&]
					\& x_1
					\\
					J_2 \ar[ur, dash]
				\end{tikzcd}
			}^2	
		}
		\\
		&=
		\de{
			\mean{
				\abs{
					\begin{tikzcd} [ampersand replacement=\&]
						J_1 \ar[dr, dash] \& 
						\\
						\& x_2
					\end{tikzcd}
				}^2	
			}
			+
			\mean{
				\abs{
					\begin{tikzcd} [ampersand replacement=\&]
						\\
						J_2 \ar[r, dash]\& x_2
					\end{tikzcd}
				}^2	
			}	
			}
		\de{
			\mean{
				\abs{
					\begin{tikzcd} [ampersand replacement=\&]
						J_1 \ar[r, dash] \& x_1 
						\\
					\end{tikzcd}
				}^2	
			}
			+
			\mean{
				\abs{
					\begin{tikzcd} [ampersand replacement=\&]
						\& x_1
						\\
						J_2 \ar[ur, dash]
					\end{tikzcd}
				}^2	
			}
			}.
			\label{IndMeasurements}	
		}
	Using the relation
	\splt{
		\mean{
			\abs{
				\begin{tikzcd} [ampersand replacement=\&]
					J_i \ar[r, dash]\& x_j
				\end{tikzcd}
				}^2
			}
			\mean{
				\abs{
					\begin{tikzcd} [ampersand replacement=\&]
						J_m \ar[r, dash]\& x_n
					\end{tikzcd}
				}^2
			}
		=
		\mean{
			\abs{
				\begin{tikzcd} [ampersand replacement=\&]
					J_i \ar[r, dash]\& x_j
					\\
					J_m \ar[r, dash]\& x_n
				\end{tikzcd}
			}^2
		},
	}
	one can rewrite (\ref{IndMeasurements}) as
	\splt{
		&\text{Tr}\de{\rho_{\text{ch}}\varphi^\dagger(x_1)\varphi(x_1)}\text{Tr}\de{\rho_{\text{ch}}\varphi^\dagger(x_2)\varphi(x_2)}
		\\
		&=
		\mean{
			\abs{
				\begin{tikzcd} [ampersand replacement=\&]
					J_1 \ar[r, dash]\& x_1
					\\
					J_2 \ar[r, dash]\& x_2
				\end{tikzcd}
			}^2
		}
		+
		\mean{
			\abs{
				\begin{tikzcd} [ampersand replacement=\&]
					J_1 \ar[dr, dash] \& x_1
					\\
					J_2\ar[ur, dash, crossing over] \& x_2
				\end{tikzcd}
			}^2
		}
		+
		\mean{
			\abs{	
				\begin{tikzcd} [ampersand replacement=\&]
					J_1 \ar[r, dash]\ar[dr, dash]\& x_1
					\\
					J_2 \& x_2
				\end{tikzcd}
			}^2
		}	
		+
		\mean{
			\abs{
				\begin{tikzcd} [ampersand replacement=\&]
					J_1 \& x_1
					\\
					J_2 \ar[r, dash]\ar[ur, dash]\& x_2
				\end{tikzcd} 
			}^2
		}.
		}
	Therefore, the only distinction between the independent measurements and the correlated one is the interference between the two indistinguishable amplitudes:
	\splt{
		&\frac{
				\text{Tr}\de{\rho_{\text{ch}} \varphi^\dagger(x)\varphi^\dagger(y) \varphi(y) \varphi(x) }
			}
			{
				\text{Tr}\de{\rho_{\text{ch}} \varphi^\dagger(x) \varphi(x) }
				\text{Tr}\de{\rho_{\text{ch}} \varphi^\dagger(y) \varphi(y) }
			}
		\\	
		&=
		1
		+
		\frac{
				\de{
					\begin{tikzcd} [ampersand replacement=\&]
						J_1 \ar[r, dash]\& x_1
						\\
						J_2 \ar[r, dash]\& x_2
					\end{tikzcd}
				}
				\de{
					\begin{tikzcd} [ampersand replacement=\&]
						J_1 \ar[dr, dash] \& x_1
						\\
						J_2\ar[ur, dash, crossing over] \& x_2
					\end{tikzcd}
				}^*
				+
				\de{
					\begin{tikzcd} [ampersand replacement=\&]
						J_1 \ar[r, dash]\& x_1
						\\
						J_2 \ar[r, dash]\& x_2
					\end{tikzcd}
				}^*
				\de{
					\begin{tikzcd} [ampersand replacement=\&]
						J_1 \ar[dr, dash] \& x_1
						\\
						J_2\ar[ur, dash, crossing over] \& x_2
					\end{tikzcd}
				}
			}
			{
				\mean{
					\abs{
						\begin{tikzcd} [ampersand replacement=\&]
							J_1 \ar[r, dash]\& x_1
							\\
							J_2 \ar[r, dash]\& x_2
						\end{tikzcd}
					}^2
				}
				+
				\mean{
					\abs{
						\begin{tikzcd} [ampersand replacement=\&]
							J_1 \ar[dr, dash] \& x_1
							\\
							J_2\ar[ur, dash, crossing over] \& x_2
						\end{tikzcd}
					}^2
				}
				+
				\mean{
					\abs{	
						\begin{tikzcd} [ampersand replacement=\&]
							J_1 \ar[r, dash]\ar[dr, dash]\& x_1
							\\
							J_2 \& x_2
						\end{tikzcd}
					}^2
				}	
				+
				\mean{
					\abs{
						\begin{tikzcd} [ampersand replacement=\&]
							J_1 \& x_1
							\\
							J_2 \ar[r, dash]\ar[ur, dash]\& x_2
						\end{tikzcd} 
					}^2
				}
			}.
		}
	Not only is the correlation a result of standard quantum mechanical wave function interference, it's also a result of a two particle amplitude interference. The coherent case, on the other hand won't have correlations because of the lack of destructive interference between amplitudes. To see that, just note that the expression for independent measurements
	\splt{
		&\text{Tr}\de{\rho_{\text{co}} \varphi^\dagger\de{x_1}\varphi\de{x_1} }
		\text{Tr}\de{\rho_{\text{co}} \varphi^\dagger\de{x_2}\varphi\de{x_2} }
		\\
		&=
			\abs{
				\begin{tikzcd} [ampersand replacement=\&]
					J_1 \ar[dr, dash] \& 
					\\
					\& x_2
				\end{tikzcd}
				+
				\begin{tikzcd} [ampersand replacement=\&]
					\\
					J_2 \ar[r, dash]\& x_2
				\end{tikzcd}
			}^2	
			\abs{
				\begin{tikzcd} [ampersand replacement=\&]
					J_1 \ar[r, dash] \& x_1 
					\\
				\end{tikzcd}
				+
				\begin{tikzcd} [ampersand replacement=\&]
					\& x_1
					\\
					J_2 \ar[ur, dash]
				\end{tikzcd}
			}^2	
		}  
	can be rewritten as
	\splt{&\text{Tr}\de{\rho_{\text{co}} \varphi^\dagger\de{x_1}\varphi\de{x_1} }
		\text{Tr}\de{\rho_{\text{co}} \varphi^\dagger\de{x_2}\varphi\de{x_2} }
		\\
		&=
		\abs{
			\begin{tikzcd} [ampersand replacement=\&]
				J_1 \ar[r, dash]\& x_1
				\\
				J_2 \ar[r, dash]\& x_2
			\end{tikzcd}
			+
			\begin{tikzcd} [ampersand replacement=\&]
				J_1 \ar[dr, dash] \& x_1
				\\
				J_2\ar[ur, dash, crossing over] \& x_2
			\end{tikzcd}
			+
			\begin{tikzcd} [ampersand replacement=\&]
				J_1 \ar[r, dash]\ar[dr, dash]\& x_1
				\\
				J_2 \& x_2
			\end{tikzcd}
			+
			\begin{tikzcd} [ampersand replacement=\&]
				J_1 \& x_1
				\\
				J_2 \ar[r, dash]\ar[ur, dash]\& x_2
			\end{tikzcd} 
		}^2.
		}
	Which is the same as the joint measurements. What if we have a very large number of sub-sources? As we saw before, in this situation the single sub-source double emissions
	\splt{
		\begin{tikzcd} [ampersand replacement=\&]
			J_i \ar[r, dash]\ar[dr, dash]\& x_1
			\\
			\& x_2
		\end{tikzcd}
		}
	are negligible. In this situation, no two different pairs of emissions interfere, the correlation comes from the constructive interference between each emission pair amplitude:
	\splt{
			\text{Tr}\de{\rho_{\text{ch}} \varphi^\dagger(x)\varphi^\dagger(y) \varphi(y) \varphi(x) }
			=
			\frac{1}{2}
			\sum\limits_{i \not = j} 
			\mean{
				\abs{
					\begin{tikzcd} [ampersand replacement=\&]
						J_i \ar[r, dash]\& x_1
						\\
						J_j \ar[r, dash]\& x_2
					\end{tikzcd}
					+
					\begin{tikzcd} [ampersand replacement=\&]
						J_i \ar[dr, dash] \& x_1
						\\
						J_j\ar[ur, dash, crossing over] \& x_2
					\end{tikzcd}
					}^2
				}.
		}
	There are no interferences between different pairs of sub-sources. That is,
	\splt{
		\mean{
			\de{
			\begin{tikzcd} [ampersand replacement=\&]
				J_i \ar[r, dash]\& x_1
				\\
				J_j \ar[r, dash]\& x_2
			\end{tikzcd}
			+
			\begin{tikzcd} [ampersand replacement=\&]
				J_i \ar[dr, dash] \& x_1
				\\
				J_j\ar[ur, dash, crossing over] \& x_2
			\end{tikzcd}
			}
			\de{
			\begin{tikzcd} [ampersand replacement=\&]
				J_m \ar[r, dash]\& x_1
				\\
				J_n \ar[r, dash]\& x_2
			\end{tikzcd}
			+
			\begin{tikzcd} [ampersand replacement=\&]
				J_m \ar[dr, dash] \& x_1
				\\
				J_n\ar[ur, dash, crossing over] \& x_2
			\end{tikzcd}
			}^*
		}
		=
		0,
	}
	for any two different pairs $\DE{i,j}$ and $\DE{m,n}$ of sub-sources. Therefore, the HBT effect or correlations is really the result of the symmetrization of the two particle wave function when we have chaotic sources. Correlations which are a result of the symmetrization of the wave function are usually called Bose-Einstein correlations.
	
	To answer the question of whether the HBT effect is a quantum effect or not, we will take the position that it can be a quantum effect. There are certain situations such as the classic HBT experiment where a classical explanation can be given, but that doesn't seem to be always possible\cite{Shih2020,Scarcelli2006}. With that out of the way, one must also keep in mind, although the effect is quantum mechanical for chaotic sources in our opinion, classical states of the field can exhibit correlations, just take a stochastic classical field as an example. In spite of the quantum interpretation we have given, for a very large number of sources, a classical stochastic field would be able to explain the correlation resulting from a chaotic source. So, it's not clear whether one could deduce quantum properties of the field from positive correlations such as the one we found for a chaotic field, these can always be explained by classical fields\cite{Ballentine2014,Baym1990}. The only kind of correlations that can't be explained by classical fields are the negative ones\cite{Mandel1986,Paul1982,Ballentine2014}.  In fact, HBT has recently been argued to be analogous to these \cite{hbtquantum}
	
        \bibliography{hbtcosmo}
\end{document}